\documentclass[
  nojss,
  shortnames,
  noheadings]{jss}

\usepackage[utf8]{inputenc}

\author{
James A. Scott\\Imperial College London \And Axel Gandy\\Imperial College
London \And Swapnil Mishra\\Imperial College London \AND Samir
Bhatt\\University of Copenhagen \And Seth Flaxman\\Imperial College
London \And H. Juliette T. Unwin\\Imperial College London \AND Jonathan
Ish-Horowicz\\Imperial College London
}
\title{\pkg{epidemia}: An \proglang{R} Package for Bayesian, Semi-Mechanistic Modeling of Infectious Diseases}

\Plainauthor{James A. Scott, Axel Gandy, Swapnil Mishra, Samir Bhatt, Seth
Flaxman, H. Juliette T. Unwin, Jonathan Ish-Horowicz}
\Plaintitle{epidemia: An R Package for Semi-Mechanistic Bayesian
Modelling of Infectious Diseases using Point Processes.}
\Shorttitle{\pkg{epidemia}: Bayesian Modeling of Infectious Diseases}

\Abstract{
This article introduces \pkg{epidemia}, an \proglang{R} package for
Bayesian, regression-oriented modeling of infectious diseases. The
implemented models define a likelihood for all observed data while also
explicitly modeling transmission dynamics: an approach often termed as
\textit{semi-mechanistic}. Infections are propagated over time using
renewal equations. This approach is inspired by self-exciting,
continuous-time point processes such as the Hawkes process. A variety of
inferential tasks can be performed using the package. Key
epidemiological quantities, including reproduction numbers and latent
infections, may be estimated within the framework. The models may be
used to evaluate the determinants of changes in transmission rates,
including the effects of control measures. Epidemic dynamics may be
simulated either from a fitted model or a ``prior'' model; allowing for
prior/posterior predictive checks, experimentation, and forecasting.
}

\Plainkeywords{epidemiology, infectious diseases, hierarchical
models, hawkes processes, bayesian}

\Address{
    James Scott\\
    Imperial College London\\
    Department of Mathematics\\
526 Huxley Building\\
London, United Kingdom\\
SW2 7AZ\\
  E-mail: \email{james.scott15@imperial.ac.uk}\\
  
              }

\usepackage{amsmath}
\usepackage{booktabs}
\usepackage{longtable}
\usepackage{array}
\usepackage{multirow}
\usepackage{wrapfig}
\usepackage{float}
\usepackage{pdflscape}
\usepackage{tabu}
\usepackage{threeparttable}
\usepackage{threeparttablex}
\usepackage[normalem]{ulem}
\usepackage[utf8]{inputenc}
\usepackage{makecell}
\usepackage{xcolor}
\usepackage{tikz}
\usepackage{adjustbox}
\usepackage{layouts}
\usepackage{subcaption}
\usepackage{times}
\usepackage{setspace}
\usepackage{standalone}

\usetikzlibrary{shapes,arrows}
\definecolor{airforceblue}{rgb}{0.36, 0.54, 0.66}
\definecolor{amaranth}{rgb}{0.9, 0.17, 0.31}
\definecolor{ao(english)}{rgb}{0.0, 0.5, 0.0}
\definecolor{aquamarine}{rgb}{0.5, 1.0, 0.83}
\definecolor{anti-flashwhite}{rgb}{0.95, 0.95, 0.96}
\definecolor{arsenic}{rgb}{0.23, 0.27, 0.29}
\definecolor{amber}{rgb}{1.0, 0.49, 0.0}

\tikzstyle{tparam} = [rectangle, draw = amaranth, line width = 0.5mm, minimum size = 20pt, inner sep=5pt, align = center, rounded corners=.2cm, 
minimum width=2.4cm, fill=white]
\tikzstyle{dummy} = [minimum size = 20pt, minimum height=30pt, minimum width=2.4cm]
\tikzstyle{param} = [tparam, draw = ao(english)]
\tikzstyle{quant} = [tparam, draw = gray, fill = gray!20]
\tikzstyle{obs} = [tparam, draw = airforceblue]
\tikzstyle{line} = [draw, -latex', ultra thick, color=gray!80]

\begin{document}

\hypertarget{introduction}{%
\section{Introduction}\label{introduction}}

The open-source R \citep{rcore_2011} package \pkg{epidemia} provides a
framework for Bayesian, regression-oriented modeling of the temporal
dynamics of infectious diseases. Typically, but not exclusively, these
models are fit to areal time-series; i.e.~aggregated event counts for a
given population and period. Disease dynamics are described explicitly;
observed data are linked to latent infections, which are in turn modeled
as a self-exciting process tempered by time-varying reproduction
numbers. Regression models are specified for several objects in the
model. For example, reproduction numbers are expressed as a transformed
predictor, which may include both covariates and autoregressive terms. A
range of prior distributions can be specified for unknown parameters by
leveraging the functionality of \pkg{rstanarm} \citep{goodrich_2020}.
Multilevel models are supported by partially pooling covariate effects
appearing in the predictor for reproduction numbers between multiple
populations.

The mathematical framework motivating the implemented models has been
described in \citet{Bhatt2020}. Specific analyses using such models have
appeared during the COVID-19 pandemic, and have been used to estimate
the effect of control measures
\citep{Flaxman2020, Mellan_2020, Olney_2021}, and to forecast disease
dynamics under assumed epidemiological parameters and mitigation
scenarios \citep{Vollmer_2020, Hawryluk_2020}. The modeling approach has
been extended to estimate differences in transmissibility between
COVID-19 lineages \citep{Faria_2021, Volz_2021}.

Models of infectious disease dynamics are commonly classified as either
mechanistic or statistical \citep{Myers2000}. Mechanistic models derive
infection dynamics from theoretical considerations over how diseases
spread within and between communities. An example of this are
deterministic compartmental models (DCMs)
\citep{kermack_1927, kermack_1932, kermack_1933}, which propose
differential equations that govern the change in infections over time.
These equations are motivated by contacts between individuals in
susceptible and infected classes. Purely statistical models, on the
other hand, make few assumptions over the transmission mechanism, and
instead infer future dynamics from the history of the process and
related covariates. Examples include Generalized Linear Models (GLMs),
time series approaches including Auto Regressive Integrated Moving
Average (ARIMA) \citep{box_1962}, and more modern forecasting methods
based on machine learning.

\pkg{epidemia} provides models which are \textit{semi-mechanistic}.
These are statistical models that explicitly describe infection
dynamics. Self-exciting processes are used to propagate infections in
discrete time. Previous infections directly precipatate new infections.
Moreover, the memory kernel of the process allows an individual's
infectiousness to depend explicitly on the time since infection. This
approach has been used in multiple previous works
\citep{fraser_2007, Cori2013, nouvellet_2018, cauchemez_2008} and has
been shown to correspond to a Susceptible-Exposed-Infected-Recovered
(SEIR) model when a particular form for the generation distribution is
used \citep{champredon_2018}. In addition, population adjustments may be
applied to account for depletion of the susceptible population. The
models are \textit{statistical} in the sense that they define a
likelihood function for the observed data. After also specifying prior
distributions for model parameters, samples from the posterior can then
be obtained using either Hamiltonian Monte Carlo or Variational Bayes
methods.

The Bayesian approach has certain advantages in this context. Several
aspects of these models are fundamentally unidentified
\citep{Roosa_2019}. For most diseases, infection counts are not fully
observable and suffer from under-reporting \citep{Gibbons_2014}.
Recorded counts could be explained by a high infection and low
ascertainment regime, or alternatively by low infections and high
ascertainment. If a series of mitigation efforts are applied in sequence
to control an epidemic, then the effects may be confounded and difficult
to disentangle \citep{Bhatt2020}. Bayesian approaches using MCMC allow
full exploration of posterior correlations between such coupled
parameters. Informative, or weakly informative, priors may be
incorporated to regularize, and help to mitigate identifiability
problems, which may otherwise pose difficulties for sampling
\citep{Gelman_2008, Gelman_2013}.

\pkg{epidemia}'s functionality can be used for a number of purposes. A
researcher can simulate infection dynamics under assumed parameters by
setting tight priors around the assumed values. It is then possible to
sample directly from the prior distribution without conditioning on
data. This allows \textit{in-silico} experimentation; for example, to
assess the effect of varying a single parameter (reproduction numbers,
seeded infections, incubation period). Another goal of modeling is to
assess whether a simple and parsimonious model of reality can replicate
observed phenomena. This helps to isolate processes helpful for
explaining the data. Models of varying complexity can be specified
within \pkg{epidemia}, largely as a result of it's regression-oriented
framework. Posterior predictive checks can be used to assess model fit.
If the model is deemed misspecified, additional features may be
considered. This could be modeling population adjustments, explicit
modeling of super-spreader events \citep{Wong_2020}, alternative and
over-dispersed models for the data, or more flexible functional forms
for reproduction numbers or ascertainment rates. This can be done
rapidly within \pkg{epidemia}'s framework.

Forecasting models are critical during an ongoing epidemic as they are
used to inform policy decisions under uncertainty. As a sign of their
importance, the United States Centers for Disease Control and Prevention
(CDC) has run a series of forecasting challenges, including the FluSight
seasonal forecasting challenges since 2015
(\url{https://www.cdc.gov/flu/weekly/flusight/}) and more recently the
Covid-19 Forecast hub (\url{https://covid19forecasthub.org/}). Similar
challenges have been run by the European Center for Disease Prevention
and Control (ECDC) (\url{https://covid19forecasthub.eu/}). Long-term
forecasts quantify the cost of an unmitigated epidemic, and provide a
baseline from which to infer the effects of control measures. Short-term
forecasts are crucial in informing decisions on how to distribute
resources such as PPE or respirators, or whether hospitals should
increase capacity and cancel less urgent procedures. Traditional
statistical approaches often give unrealistic long-term forecasts as
they do not explicitly account for population effects. The
semi-mechanistic approach of \pkg{epidemia} combines the strengths of
statistical approaches with plausible infection dynamics, and can thus
be used for forecasting at different tenures.

The rest of this article is organized as follows. Section
\ref{sec:relatedpackages} discusses alternative \proglang{R} packages
for epidemiology, and highlights the unique features of \pkg{epidemia}.
Section \ref{sec:modeldescription} introduces the basic model and
various extensions. Sections \ref{sec:installation} and
\ref{sec:modelimplementation} provide installation instructions and
introduce some of the main functions required to specify and fit the
models. We proceed in Section \ref{sec:epi_examples} to demonstrate usage of
the package on two examples. The first example considers the task of
inferring time-varying reproduction numbers, while the second attempts
to infer the effects of control measures using a multilevel model.
Finally, we conclude in Section \ref{sec:conclusions}.

\hypertarget{sec:relatedpackages}{%
\subsection{Related packages}\label{sec:relatedpackages}}

The Comprehensive R Archive Network (CRAN)
(\url{https://cran.r-project.org/}) provides a rich ecosystem of
\proglang{R} packages dedicated to epidemiological analysis. The R
Epidemics Consortium website
(\url{https://www.repidemicsconsortium.org/}) lists a number of these.
Packages that model infectious disease dynamics vary significantly by
the methods used to model transmission. \pkg{RLadyBug}
\citep{hohle_2007} is a \proglang{R} package for parameter estimation
and simulation for stochastic compartmental models, including SEIR-type
models. Both likelihood-based and Bayesian inference are supported.
\pkg{amei} \citep{merl_2010} provides online inference for a stochastic
SIR model with a negative-binomial transmission function, however the
primary focus is on identifying optimal intervention strategies. See
\citet{Anderson_2000} for an introduction to stochastic epidemic
modeling.

\pkg{epinet} \citep{Groendyke_2018} and \pkg{epimodel}
\citep{Jenness_2018} provide functionality to simulate compartmental
models over contact networks. \pkg{epinet} uses the class of
dyadic-independent exponential random graph models (ERGMs) to model the
network, and perform full Bayesian inference over model parameters.
\pkg{epimodel} considers instead dynamic networks, inferring only
network parameters and assuming epidemic parameters to be known.

Epidemic data often presents in the form of areal data, recording event
counts over disjoint groups during discrete time intervals. This is the
prototypical data type supported within \pkg{epidemia}. Areal data can
be modeled using purely statistical methods. The \code{glm()} function
in \pkg{stats} can be used to fit simple time-series models to count
data. The package \pkg{acp} \citep{Vasileios_2015} allows for fitting
autoregressive Poisson regression (ACP) models to count data, with
potentially additional covariates. \pkg{tscount} \citep{Liboschik_2017}
expands on \pkg{acp}, in particularly providing more flexible link
functions and over-dispersed distributions.

Like \pkg{epidemia}, the \proglang{R} package \pkg{Surveillance}
\citep{Meyer_2017} implements regression-oriented modeling of epidemic
dynamics. The package offers models for three different spatial and
temporal resolutions of epidemic data. For areal data, which is the
focus of \pkg{epidemia}, the authors implement a multivariate
time-series approach \citep{Held_2005, Paul_2008, Paul_2011, Held_2012}.
This model differs from the semi-mechanistic approach used here in
several ways. First, the model has no mechanistic component: neither
infections and transmission are explicitly described. The model is
similar in form to a vector autoregressive model of order 1
(\(\text{VAR}(1)\)). The lag 1 assumption implies that each count series
is Markovian. In \pkg{epidemia}, the infection process has an
interpretation as an AR process with both order and coefficients
determined by the generation distribution. This can therefore model more
flexible temporal dependence in observed data.

\pkg{EpiEstim} \citep{Cori2013, Cori2020} infers time-varying
reproduction numbers \(R_t\) using case counts over time and an
approximation of the disease's generation distribution. Infection
incidence is assumed to follow a Poisson process with expectation given
by a renewal equation. \pkg{R0} \citep{obadia_2012} implements
techniques for estimating both initial and time-varying transmission
rates. In particular, the package implements the method of
\citet{wallinga_2004}, which bases estimates off a probabilistic
reconstruction of transmission trees. \pkg{epidemia} differs from these
packages in several ways. First, if infection counts are low then the
Poisson assumption may be too restrictive, as super-spreader events can
lead to over-dispersion in the infection process. Our framework permits
over-dispersed distributions for modeling latent infections. Second,
\pkg{epidemia} allows flexible prior models for \(R_t\), including the
ability to use time-series methods. For example, \(R_t\) can be
parameterized as a random walk. Finally, infections over time are often
unobserved, and subject to under-reporting that is both space and time
dependent. We account for this by providing flexible observation models
motivated by survival processes. Several count data series may be used
simultaneously within the model in order to leverage additional
information on \(R_t\).

The probabilistic programming language \proglang{Stan} \citep{stan_2020}
has been used extensively to specify and fit Bayesian models for disease
transmission during the Covid-19 pandemic. Examples analyses include
\citet{Flaxman2020}, \citet{Hauser_2020} and \citet{Doremalen_2020}. For
tutorials on implementing such models, see for example
\citet{Grinsztajn_2021} or \citet{Chatzilena_2019}. \pkg{epidemia} uses
the framework offered by \proglang{Stan} to both specify and fit models.
User-specified models are internally translated into data that is passed
to a precompiled \proglang{Stan} program. The models are fit using
sampling methods from \pkg{rstan} \citep{rstan_2020}.

\hypertarget{sec:modeldescription}{%
\section{Model Description}\label{sec:modeldescription}}

Here, we present the modeling framework implemented by the package.
Section \ref{sec:basic_model} outlines the bare-bones version of the
model, which is elaborated on in Sections \ref{sec:epi_observations},
\ref{sec:epi_infections} and \ref{sec:transmission}. Section
\ref{sec:extensions} extends the model and introduces multilevel
modeling, treating infections as parameters, and accounting for
population effects.

\hypertarget{sec:basic_model}{%
\subsection{Basic Model}\label{sec:basic_model}}

We now formulate the basic version of the model for one homogeneous
population. The same model can be used for multiple regions or groups
jointly. Suppose we observe a non-negative time series of count data
\(Y = (Y_1, \ldots Y_n)\) for a single population. This could for
example be daily death or case incidence. \(Y_t\) is modeled as deriving
from past new infections \(i_s\), \(s < t\), and some parameter
\(\alpha_t > 0\), a multiplier, which in most contexts represents an
instantaneous \emph{ascertainment rate}. The general model can be
expressed as \begin{align} 
Y_t & \sim p(y_t , \phi), \label{eq:ysampling} \\
y_t & = \alpha_t \sum_{s < t} i_s \pi_{t-s} \label{eq:obsmean}, 
\end{align} where \(y_t\) is the expected value of the data distribution
and \(\phi\) is an auxiliary parameter. \(\pi_{k}\) is typically the
time distribution from an infection to an observation, which we refer to
as the \emph{infection to observation} distribution. More generally,
however, \(\pi_k\) can be used to obtain any linear combination of past
infections. New infections \(i_t\) at times \(t>0\) are modeled through
a renewal equation, and are tempered by a non-negative parameter \(R_t\)
which represents the reproduction number at time \(t\). Formally
\begin{align} 
i_t &= R_t \sum_{s < t} i_s g_{t-s}, \label{eq:renewal2}
\end{align} where \(g_k\) is a probability mass function for the time
between infections. The recursion is initialized with \emph{seeded}
infections \(i_{v:0}\), \(v < 0\), which are treated as unknown
parameters. All parameters are assigned priors, i.e. \begin{equation} 
i_{v:0}, R, \phi, \alpha \sim p(\cdot),
\end{equation} where \(R = (R_1, \ldots, R_n)\) and
\(\alpha = (\alpha_1, \ldots, \alpha_n)\). The posterior distribution is
then proportional to prior and likelihood, i.e.~ \begin{equation}
p(i_{v:0}, R, \phi, \alpha \mid Y) \propto p(i_{v:0})p(R)p(\phi)p(\alpha) \prod_{t} p(Y_t \mid y_t, \phi). 
\end{equation} This posterior distribution is represented in a
\proglang{Stan} program, and an adaptive Hamiltonian Monte Carlo sampler
\citep{hoffman_2014} is used to approximately draw samples from it.
These samples allow for inference on the parameters, in addition to
simulating data from the posterior predictive distribution.

Reproduction numbers \(R\) and multipliers \(\alpha\) can be modeled
flexibly with Bayesian regression models, and by sharing parameters, are
the means by which multiple regions or groups are tied together through
multilevel models. One can, for example, model \(R\) as depending on a
binary covariate for a control measure, say full lockdown. The
coefficient for this can be \emph{partially pooled} between multiple
populations. The effect is to share information between groups, while
still permitting between group variation.

\hypertarget{sec:epi_observations}{%
\subsection{Observations}\label{sec:epi_observations}}

As mentioned, \(Y_t\) is usually a count of some event type occurring at
time \(t\). These events are precipitated by past infections.
Prototypical examples include daily cases or deaths. \(\alpha_t\) is a
multiplier, and when modeling count data, it typically is interpreted as
an \emph{ascertainment rate}, i.e.~the proportion of events at time
\(t\) that are recorded in the data. For case or death data this would
be the infection ascertainment rate (IAR) or the infection fatality rate
(IFR) respectively.

The multiplier \(\alpha\) plays a similar role for observations as \(R\)
does for infections; tempering expected observations for time-specific
considerations. As such, \pkg{epidemia} treats \(\alpha\) in a similar
manner to reproductions number, and allows the user to specify a
regression model for it. Section \ref{sec:transmission} discusses this
in detail in the context of reproduction numbers, and this discussion is
not repeated here. Figure \ref{fig:obs_model} in Appendix
\ref{sec:modelschematic} details the model for \(\alpha\), as well as
for observational models in general.

The sampling distribution \(p(y_t, \phi)\) (Equation
\eqref{eq:ysampling}) should generally be informed by parts of the data
generating mechanism not captured by the mean \(y_t\): i.e.~any
mechanisms which may induce additional variation around \(y_t\). Options
for \(p(y_t, \phi)\) include the Poisson, quasi-Poisson and
negative-binomial families. The Poisson family has no auxiliary
parameter \(\phi\), while for the latter two families this represents a
non-negative \emph{dispersion parameter} which is assigned a prior.

\pkg{epidemia} allows simultaneous modeling of multiple observation
vectors. In this case, we simply superscript \(Y_t^{(l)}\),
\(\alpha_{t}^{(l)}\) and \(\pi^{(l)}\), and assign independent sampling
distributions for each type. Separate regression models are then
specified for each multiplier \(\alpha_{t}^{(l)}\). Leveraging multiple
observation types can often enhance a model. For example, high quality
death data existed during the first wave of the Covid-19 pandemic in
Europe. Case data gradually increased in reliability over time, and has
the advantage of picking up changes in transmission dynamics much
quicker than death data.

\hypertarget{sec:epi_infections}{%
\subsection{Infections}\label{sec:epi_infections}}

Infections \(i_t\) propagate over time through the discrete renewal
equation \eqref{eq:renewal2}. This is \emph{self-exciting}: past
infections give rise to new infections. The theoretical motivation for
this lies in counting processes and is explained in more detail in
\citet{Bhatt2020}. The equation is connected to Hawkes processes and the
Bellman Harris branching process
\citep{bellman_1948, bellman_1952, mishra2020}. Such processes have been
used in numerous previous studies
\citep{fraser_2007, Cori2013, nouvellet_2018, cauchemez_2008}, and are
also connected to compartmental models such as the SEIR model
\citep{champredon_2018}.

Equation \eqref{eq:renewal2} implies that infections \(i_t\), \(t > 0\)
are deterministic given \(R\) and seeded infections \(i_{v:0}\).
\pkg{epidemia} sets a prior on \(i_{v:0}\) by first assuming that daily
seeds are constant over the seeding period. Formally, \(i_{k} = i\) for
each \(k \in \{v,\ldots 0\}\). The parameter \(i\) can be assigned a
range of prior distributions. One option is to model it hierarchically;
for example as \begin{align} 
i & \sim \text{Exp}(\tau^{-1}), \label{eq:seeds}\\
\tau & \sim \text{Exp}(\lambda_0) \label{eq:tau},
\end{align} where \(\lambda_0 > 0\) is a rate hyperparameter. This prior
is uninformative, and allows seeds to be largely determined by initial
transmission rates and the chosen start date of the epidemic.

Several extensions to the infection model are possible in
\pkg{epidemia}, including extending \eqref{eq:renewal2} to better capture
dynamics such as super-spreading events, and also adjusting the process
for the size of the remaining susceptible population. These extensions
are discussed in Section \ref{sec:latent} and \ref{sec:epi_popadjust}
respectively. The basic infection model is shown in Figure
\ref{fig:inf_basic} in Appendix \ref{sec:modelschematic}.

\hypertarget{sec:transmission}{%
\subsection{Transmission}\label{sec:transmission}}

Reproduction numbers are modeled flexibly. One can form a linear
predictor consisting of fixed effects, random effects and
autocorrelation terms, which is then transformed via a suitable link
function. Formally \begin{equation} 
R = g^{-1}(\eta),
\end{equation} where \(g\) is a link function and \(\eta\) is a linear
predictor. In full generality, \(\eta\) can be expressed as
\begin{equation} 
\eta = \beta_0 + X \beta + Z b + Q \gamma, \label{eq:linpred}
\end{equation} where \(X\) is an \(n \times p\) model matrix, \(Z\) is
an \(n \times q\) model matrix for the \(q\)-vector of group-specific
parameters \(b\). \(Q\) is an \(n \times r\) model matrix for the
\(r\)-vector of autocorrelation terms. The columns of \(X\) are
predictors explaining changes in transmission. These could, for example,
be binary vectors encoding non-pharmaceutical interventions, as in
\citet{Flaxman2020}. A number of families can be used for the prior on
\(\beta\), including normal, cauchy, and hierarchical shrinkage
families. The parameters \(b\) are modeled hierarchically as
\begin{equation} 
b \sim N(0, \Sigma), 
\end{equation} where \(\Sigma\) is a covariance matrix that is itself
assigned a prior. The particular form for \(\Sigma\), as well as its
prior is discussed in more detail in Appendix \ref{sec:priorscov}. These
partially-pooled parameters are particularly useful when multiple
regions are being modeled simultaneously. In this case, they allow
information on transmission rates to be shared between groups.

\(Q\) is a binary matrix specifying which of the autocorrelation terms
in \(\gamma\) to include for each period \(t\). Currently,
\pkg{epidemia} supports only random walk processes. However multiple
such processes can be included, and can have increments that occur at a
different time scale to \(R\); for example weekly increments can be
used.

\hypertarget{sec:links}{%
\subsubsection{Link Functions}\label{sec:links}}

Choosing an appropriate link function \(g\) is difficult. \(R_t\) is
non-negative, but is clearly not able to grow exponentially: regardless
of the value of the linear predictor \(\eta_t\), one expects \(R_t\) to
be bounded by some maximum value \(K\). In other words, \(R_t\) has some
\emph{carrying capacity}. One of the simplest options for \(g\) is the
log-link. This satisfies non-negativity, and also allows for easily
interpretable effect sizes; a one unit change in a predictor scales
\(R_t\) by a constant factor. Nonetheless, it does not respect the carry
capacity \(K\), often placing too much prior mass on large values of
\(R_t\). With this in mind, \pkg{epidemia} offers an alternative link
function satisfying \begin{equation}
g^{-1}(x) = \frac{K}{1 + e^{-x}} \label{eq:scaledlogit}. 
\end{equation} This is a generalization of the logit-link, and we refer
to it as the \emph{scaled-logit}.

\hypertarget{sec:extensions}{%
\subsection{Extensions}\label{sec:extensions}}

Various extensions to the basic model just presented are possible,
including multilevel modeling, adding variation to the infection
process, and explicitly accounting for population effects. These are
discussed in turn.

\hypertarget{sec:mult}{%
\subsubsection{Joint Modeling of Multiple Populations}\label{sec:mult}}

Consider modeling the evolution of an epidemic across multiple regions
or populations. Of course, separate models can be specified for each
group. This approach is fast as each model can be fit in parallel.
Nonetheless, often there is little high quality data for some groups,
particularly in the early stages of an epidemic. A joint model can
benefit from improved parameter estimation by \emph{sharing signal
across groups}. This can be done by partially or fully pooling effects
underlying reproduction numbers \(R\).

We give an example for concreteness. Suppose the task is to infer the
effect of a series of \(p\) control measures on transmission rates.
Letting \(R^{(m)}\) be the vector of reproduction numbers for the
\(m\)\textsuperscript{th} group, one could write \begin{equation}
R^{(m)} = g^{-1}\left( \beta_0 + b_0^{(m)} + X^{(m)}
(\beta + b^{(m)}) \right), 
\end{equation} where \(X^{(m)}\) is an \(n \times p\) matrix whose rows
are binary vectors indicating which of the \(p\) measures have been
implemented in the \(m\)\textsuperscript{th} group at that point in
time. The parameters \(b_{0}^{(m)}\) allow each region to have its own
initial reproduction number \(R_0\), while \(b^{(m)}\) allow for
region-specific policy effects. These parameters can be partially pooled
by letting \begin{equation} 
(b_0^{(m)}, b^{(m)}) \sim N(0,\tilde{\Sigma}),
\end{equation} for each \(m\), and assigning a hyperprior to the
covariance matrix \(\tilde{\Sigma}\).

In addition to hierarchical modeling of parameters making up \(R\),
seeded infections are also modeled hierarchically. Equations
\eqref{eq:seeds} and \eqref{eq:tau} are replaced with \begin{align}
i^{(m)} &\sim \text{Exp}(\tau^{-1}), \\
\tau & \sim \text{Exp}(\lambda_0),
\end{align} where \(i^{(m)}\) is the daily seeded infections for the
\(m\)\textsuperscript{th} group.

\hypertarget{sec:latent}{%
\subsubsection{Infections as Parameters}\label{sec:latent}}

Recall the renewal equation (Equation \eqref{eq:renewal2}) which
describes how infections propagate in the basic model. Infections
\(i_t\) for \(t > 0\) are a deterministic function of seeds \(i_{v:0}\)
and reproduction numbers \(R\). If infections counts are large, then
this process may be realistic enough. However, when infection counts are
low, there could variation in day-to-day infections caused by a heavy
tailed offspring distribution and super-spreader events. This may cause
actual infections to deviate from those implied by the renewal equation.
Although the \emph{expected} number of offspring of any given infection
is driven by \(R\), in practice the actual number of offspring can
exhibit considerable variation around this. To capture this randomness,
replace Equation \eqref{eq:renewal2} with \begin{align} 
i_t &\sim p(i_t', d), \label{eq:infextended} \\
i_{t}' &= R_t \sum_{s < t} i_s g_{t-s}.
\end{align} This treats \(i_t\) as latent parameters which must be
sampled. Instead, the \textit{mean value} is described by the renewal
equation. \(p(i_t', d)\) is parameterised by the mean and the
coefficient of dispersion \(d\), which is assigned a prior. This
extension can be motivated formally through counting processes. Please
see \citet{Bhatt2020} for more details.

\hypertarget{sec:epi_popadjust}{%
\subsubsection{Depletion of the Susceptible
Population}\label{sec:epi_popadjust}}

Nothing in Equation \eqref{eq:renewal2} prevents cumulative infections
from exceeding the total population size \(P\). In particular if
\(R_t > 1\) then infections can grow exponentially over time. This does
not always present a problem for modeling. Indeed the posterior
distribution usually constrains past infections to reasonable values.
Nonetheless, forecasting in the basic model will be unrealistic if
projected infections grow too large. As the susceptible population
diminishes, the transmission rate is expected to fall.

\pkg{epidemia} can apply a simple transformation to ensure that
cumulative infections remain bounded by \(P\), and that transmission
rates are adjusted for changes in the susceptible population. Let
\(S_t \in [0,P]\) be the number of susceptible individuals in the
population at time \(t\). Just like infections, this is treated as a
continuous quantity. \(S_t\) consists of those who have not been
infected by time \(t\), and have not been removed from the susceptible
class by other means; i.e.~vaccination.

Let \(i'_t\) denote \textit{unadjusted infections} from the model. This
is given by \eqref{eq:renewal2} in the basic model or by
\eqref{eq:infextended} if the extension of Section \ref{sec:latent} is
applied. These are interpreted as the number of infections if the entire
population were susceptible. These are adjusted with

\begin{equation} \label{eq:pop_adjust}
i_t = S_{t-1}\left( 1 - \exp\left(-\frac{i'_t}{P}\right)\right). 
\end{equation}

The motivation for this is provided in \citet{Bhatt2020}. Equation
\eqref{eq:pop_adjust} satisfies intuitive properties: if \(i'_t = 0\)
then \(i_t = 0\), and as \(i'_t \to \infty\) we have that
\(i_t \to S_{t-1}\). All infections at time \(t\) are then removed from
the susceptible population, so that \begin{equation}
\label{eq:st_rec} S_t = S_{t-1} - i_t
\end{equation} We are left to define \(S_{v-1}\), the susceptible
population the day before modeling begins. If this is the start of an
epidemic, it is natural to take \(S_{v-1} = P\). Nonetheless, it is
often of interest to begin modeling later, when a degree of immunity
already exists exists within the population. In this case,
\pkg{epidemia} allows the user to assign a prior distribution to
\(S_{v-1} / P\). This must lie between \(0\) and \(1\).

\hypertarget{accounting-for-vaccinations}{%
\paragraph{Accounting for
Vaccinations}\label{accounting-for-vaccinations}}

Previous infection is one avenue through which individuals are removed
from the susceptible population. Immunity can also be incurred through
vaccination. \pkg{epidemia} provides a basic way to incorporate such
effects.

Let \(v_t\) be the proportion of the susceptible population at time
\(t\) who are removed through some means other than infection. These are
individuals who have never been infected but may have been previously
vaccinated, and their immunity is assumed to have developed at time
\(t\).

\pkg{epidemia} requires \(v_t\) to be supplied by the user. Then
\eqref{eq:st_rec} is replaced with \begin{equation}
S_t = \left(S_{t-1} - i_t\right) \left(1 - v_t \right). 
\end{equation} Of course, \(v_t\) is a difficult quantity to estimate.
It requires the user to estimate the time-lag for a jab to become
effective, and to also adjust for potentially different efficacies of
jabs and doses. Recognizing this, we allow the update \begin{equation} 
S_t = \left(S_{t-1} - i_t\right) \left(1 - v_t \xi \right),
\end{equation} where \(\xi\) is a noise term that is assigned a prior
distribution. \(\xi\) helps to account for potentially systematic biases
in calculating vaccine efficacy.

\hypertarget{sec:installation}{%
\section{Installation}\label{sec:installation}}

\textbf{epidemia} requires \proglang{R} v3.5.0 or above. The package can
be installed directly from github. However, this requires you to have a
working \proglang{C++} toolchain. To ensure that this is working, please
first install \pkg{rstan} by following these
\href{https://github.com/stan-dev/rstan/wiki/RStan-Getting-Started}{installation
instructions}.

After installing \pkg{rstan}, running

\begin{CodeChunk}
\begin{CodeInput}
R> #install.packages("devtools")
R> devtools::install_github("ImperialCollegeLondon/epidemia")
\end{CodeInput}
\end{CodeChunk}

will install the latest development version of \pkg{epidemia}. If using
windows, you can alternatively install the
\href{https://github.com/ImperialCollegeLondon/epidemia/releases/latest}{binary}.
Vignettes are not currently included in the package because they are
computationally demanding, and are best viewed online.

\hypertarget{sec:modelimplementation}{%
\section{Model Implementation}\label{sec:modelimplementation}}

Here we give a high-level overview of the workflow required for defining
and fitting a model with \pkg{epidemia}. The primary model fitting
function is \code{epim()}. This takes a model description and additional
arguments relating to the fitting algorithm, and proceeds to fit the
model using a precompiled \proglang{Stan} program. This is similar to
the workflow for fitting Bayesian regression models with \pkg{rstanarm}.
A key difference, however, is that the models fit by \pkg{epidemia} are
generally complex, and are therefore inherently more difficult to
specify. We simplify this process by taking a modular approach; models
are defined through three distinct parts: transmission, infections and
observations. These components of the model are defined with the
functions \code{epirt()}, \code{epiinf()} and \code{epiobs()}
respectively.

The package contains an example dataset \code{EuropeCovid} which
contains data on daily death counts from Covid-19 in 11 European
Countries from February through May 2020, and a set of binary indicators
of non-pharmaceutical interventions. This is used as an example
throughout.

\begin{CodeChunk}
\begin{CodeInput}
R> library(dplyr)
R> library(epidemia)
R> library(rstanarm)
R> data("EuropeCovid")
\end{CodeInput}
\end{CodeChunk}

We begin by describing \code{epim()} in more detail, and then proceed to
discuss the three modeling functions.

\hypertarget{sec:fitting}{%
\subsection{Model Fitting}\label{sec:fitting}}

\code{epim()} is the only model fitting function in \pkg{epidemia}. It
has arguments \code{rt}, \code{inf}, and \code{obs} which expect a
description of the transmission model, infection model and all
observational models respectively. Together, these fully define the
joint distribution of data and parameters. Each of these model
components are described in terms of variables that are expected to live
in a single data frame, \code{data}. This data frame must be compatible
with the model components, in the sense that \emph{it holds all
variables defined in these models}. For our example, these variables are
the following.

\begin{CodeChunk}
\begin{CodeInput}
R> data <- EuropeCovid$data
R> colnames(data)
\end{CodeInput}
\begin{CodeOutput}
[1] "country"                     
[2] "date"                        
[3] "schools_universities"        
[4] "self_isolating_if_ill"       
[5] "public_events"               
[6] "lockdown"                    
[7] "social_distancing_encouraged"
[8] "deaths"                      
[9] "pop"                         
\end{CodeOutput}
\end{CodeChunk}

The \code{data} argument is described in more detail in Section
\ref{sec:data}.

In addition to taking a model description and a data frame,
\code{epim()} has various additional arguments which specify how the
model should be fit. If \code{algorithm = "sampling"} then the model
will be fit using \proglang{Stan}'s adaptive Hamiltonian Monte Carlo
sampler \citep{hoffman_2014}. This is done internally by calling
\code{sampling()} from \pkg{rstan}. If instead this is
\code{"meanfield"} or \code{"fullrank"}, then \proglang{Stan}'s
Variational Bayes algorithms \citep{Kucukelbir_2015, Kucukelbir_2017}
are employed by calling \code{vb()} from \pkg{rstan}. Any unnamed
arguments in the call to \code{epim()} are passed directly onto the
\pkg{rstan} sampling function. \code{epim()} returns a fitted model
object of class \code{epimodel}, which contains posterior samples from
the model along with other useful objects.

In general, Hamiltonian Monte Carlo should be used for final inference.
Nonetheless, this is often computationally demanding, and Variational
Bayes can often be used fruitful for quickly iterating models. All
arguments for \code{epim()} are described in Table
\ref{tab:model-imp-epim-args}.

\begin{CodeChunk}
\begin{table}[!h]

\caption{\label{tab:model-imp-epim-args} \small Formal arguments for the model fitting function \code{epim()}. The first three arguments listed below define the model to be fitted.}
\centering
\begin{tabular}[t]{>{}l>{\raggedright\arraybackslash}p{30em}}
\toprule
Argument & Description\\
\midrule
\textbf{\code{rt}} & An object of class \code{epirt}, resulting from a call to \code{epirt()} (Section \ref{sec:imp_transmission}). This defines the model for time-varying reproduction numbers $R$. See Section \ref{sec:imp_transmission} for more details.\\
\textbf{\code{inf}} & An object of class \code{epiinf}, resulting from a call to \code{epiinf()} (Section \ref{sec:imp_infections}). This entirely defines the model for infections $i_t$.\\
\textbf{\code{obs}} & Either an object of class \code{epiobs}, or a list of such objects. Each of these define a model for an observation vector in \code{data}, and result from a call to \code{epiobs()} (Section \ref{sec:imp_observations}).Each element of the list defines a model for an observed variable.\\
\textbf{\code{data}} & A dataframe with all data required for fitting the model. This includes all observations and covariates specified in the model. See Section \ref{sec:data} for more details.\\
\textbf{\code{algorithm}} & One of \code{"sampling"}, \code{"meanfield"} or \code{"fullrank"}. This determines the \pkg{rstan} sampling function to use for fitting the model. \code{"sampling"} corresponds to HMC, while \code{"meanfield"} and \code{"fullrank"} are Variational Bayes algorithms.\\
\addlinespace
\textbf{\code{group\_subset}} & If specified, a character vector naming a subset of groups/populations to include in the model.\\
\textbf{\code{prior\_PD}} & If \code{TRUE}, parameters are sampled from their prior distributions. This is useful for prior predictive checks. Defaults to \code{FALSE}.\\
\textbf{...} & Additional arguments to pass to the \pkg{rstan} function used to fit the model. If \code{algorithm = "sampling"}, then this function is \code{sampling()}. Otherwise \code{vb()} is used.\\
\bottomrule
\end{tabular}
\end{table}

\end{CodeChunk}

\hypertarget{sec:imp_transmission}{%
\subsection{Transmission}\label{sec:imp_transmission}}

\code{epirt()} defines the model for time-varying reproduction numbers,
which was described in Section \ref{sec:transmission}. Recall that these
are modeled as a transformed linear predictor. \code{epirt()} has a
\code{formula} argument which defines the linear predictor \(\eta\), an
argument \code{link} defining the link function \code{g}, and additional
arguments to specify priors on parameters making up \(\eta\).

A general \proglang{R} formula gives a symbolic description of a model.
It takes the form \code{y ~ model}, where \code{y} is the response and
\code{model} is a collection of terms separated by the \code{+}
operator. \code{model} fully defines a linear predictor used to predict
\code{y}. In this case, the ``response'' being modeled are reproduction
numbers which are unobserved. \code{epirt()} therefore requires that the
left hand side of the formula takes the form \code{R(group, date)},
where \code{group} and \code{date} refer to variables representing the
modeled populations and dates respectively. The right hand side can
consist of fixed effects, random effects, and autocorrelation terms. For
our example, a viable call to \code{epirt()} is the following.

\begin{CodeChunk}
\begin{CodeInput}
R> rt <- epirt(formula = R(country, date) ~ 1 + lockdown + public_events,
+             link = scaled_logit(7))
\end{CodeInput}
\end{CodeChunk}

Here, two fixed effects are included which represent the effects of
implementing lockdown and banning public events. These effects are
assumed constant across countries. They could alternatively be partially
pooled by using the term \code{(lockdown + public_events | country)}.
For information on how to interpret such terms, please read Appendix
\ref{sec:partial_pooling}. Using \code{link = scaled_logit(7)} lets the
link function be the scaled logit link described by Equation
\eqref{eq:scaledlogit}, where \(K = 7\) is the maximum possible value
for reproduction numbers. For simplicity, we have omitted any prior
arguments, however these should generally be specified explicitly.
Please see Appendix \ref{sec:priors} for detailed information on how to
use priors. All arguments for \texttt{epirt()} are listed in Table
\ref{tab:model-imp-epim-args}.

\begin{CodeChunk}
\begin{table}[!h]

\caption{ \label{tab:model-imp-epirt-args}\small Formal arguments for \code{epirt()}, which defines the model for $R_t$.}
\centering
\begin{tabular}[t]{>{}l>{\raggedright\arraybackslash}p{30em}}
\toprule
Argument & Description\\
\midrule
\textbf{\code{formula}} & An object of class \code{formula} which determines the linear predictor $\eta$ for $R$. The left hand side must take the form \code{R(group, date)}, where \code{group} must be a factor vector indicating group membership (i.e. country, state, age cohort), and \code{date} must be a vector of class \code{Date}. This is syntactic sugar for the reproduction number in the given group at the give date.\\
\textbf{\code{link}} & The link function $g$. Can be \code{"log"}, \code{"identity"} or a call to \code{scaled_logit()}. Defaults to \code{"log"}.\\
\textbf{\code{center}} & If \code{TRUE}, covariates specified in \code{formula} are centered to have mean zero. All priors should then be interpreted as priors on the centered covariates.\\
\textbf{\code{prior}} & Same as in \code{stan_glm()} from \pkg{rstanarm}. Defines the prior on fixed effects $\beta$. Priors provided by \pkg{rstanarm} can be used, and additionally \code{shifted_gamma}. \textbf{Note}: if \code{autoscale = TRUE} in the call to the prior function, then automatic rescaling takes place.\\
\textbf{\code{prior\_intercept}} & Same as in \code{stan_glm()} from \pkg{rstanarm}. Prior for the regression intercept $\beta_0$ (if it exists).\\
\addlinespace
\textbf{\code{prior\_covariance}} & Same as in \code{stan_glmer()} from \code{rstanarm}. Defines the prior on the covariance matrix $\Sigma$. Only use if the \code{formula} has one or more terms  of the form \code{(x | y)}, in which case there are parameters to partially pool, i.e. $b$ has positive length.\\
\textbf{\code{...}} & Additional arguments to pass to \code{model.frame()} from \pkg{stats}.\\
\bottomrule
\end{tabular}
\end{table}

\end{CodeChunk}

\hypertarget{sec:imp_infections}{%
\subsection{Infections}\label{sec:imp_infections}}

The infection model is represented by \code{epiinf()}. In the most basic
version, this defines the distribution of the generation time of the
disease, the number of days for which to seed infections, and the prior
distribution on seeded infections. These three parameters are controlled
by the arguments \code{gen}, \code{seed_days} and \code{prior_seeds}
respectively. A possible model is the following.

\begin{CodeChunk}
\begin{CodeInput}
R> inf <- epiinf(gen = EuropeCovid$si, seed_days = 6L, 
+               prior_seeds = hexp(exponential(0.02)))
\end{CodeInput}
\end{CodeChunk}

\code{EuropeCovid$si} is a numeric vector representing the distribution
for the serial interval of Covid-19. There is an implicit assumption
that the generation time can be approximated well by the serial
interval. Seeds are modeled hierarchically, and are described by
\eqref{eq:seeds} and \eqref{eq:tau}. \(\tau\) has been assigned an
exponential prior with a mean of 50. Seeded infections are assumed to
occur over a period of 6 days.

\code{epiinf()} has additional arguments that allow the user to extend
the basic model. Using \code{latent = TRUE} replaces the renewal
equation \eqref{eq:renewal2} with Equation \eqref{eq:infextended}. Daily
infections are then treated as latent parameters that are sampled along
with other parameters. The \code{family} argument specifies the
distribution \(p(i'_t, d)\), while \code{prior_aux} defines the prior on
the coefficient of dispersion \(d\).

Recall from Section \ref{sec:epi_popadjust} that the infection process may
be modified to explicitly account for changes in infection rates as the
remaining susceptible population is depleted. In particular, the
adjustment ensures that cumulative infections never breaches the
population size. It can be employed by setting \code{pop_adjust = TRUE}
and using the \code{pop} argument to point towards a static variable in
the data frame giving the population size. All argument to
\code{epiinf()} are described in Table \ref{tab:model-imp-epiinf-args}.

\begin{CodeChunk}
\begin{table}[!h]

\caption{\label{tab:model-imp-epiinf-args}\small Formal arguments for \code{epiinf()}, which defines the infection model.}
\centering
\begin{tabular}[t]{>{}l>{\raggedright\arraybackslash}p{30em}}
\toprule
Argument & Description\\
\midrule
\textbf{\code{gen}} & A numeric vector giving the probability mass function $g_k$ for the generation time of the disease (must be a probability vector).\\
\textbf{\code{seed\_days}} & An integer giving the number of days $v + 1$ for which to seed infections. Defaults to 6L.\\
\textbf{\code{prior\_seeds}} & Prior distribution on the seed parameter $i$. Defaults to  \code{hexp(prior_aux = rstanarm::exponential(0.03))}.\\
\textbf{\code{latent}} & If \code{TRUE}, treat infections as latent parameters using the extensions described in Section \ref{sec:latent}.\\
\textbf{\code{family}} & Specifies the family for the infection distribution $p(i'_t, d)$. Only used if \code{latent = TRUE}, and defaults to \code{"normal"}.\\
\addlinespace
\textbf{\code{prior\_aux}} & Prior on the auxiliary variable $d$ of $p(i'_t,d)$. This is either the variance-to-mean ratio or the coefficient of variation, depending on the value of \code{fixed_vtm}. Only used if \code{latent = TRUE}.\\
\textbf{\code{fixed\_vtm}} & If \code{TRUE}, then $p(i'_t, d)$ has a fixed variance-to-mean ratio, i.e. variance is $\sigma^2 = d i'_t$; In this case, $d$ refers to the \textit{variance-to-mean ratio}. Id \code{FALSE} then instead standard deviation is assumed proportional to the mean, in which case $d$ is the \textit{coefficient of variation}. Only used if \code{latent = TRUE}.\\
\textbf{\code{pop\_adjust}} & If \code{TRUE}, applies the population adjustment \eqref{eq:pop_adjust} to the infection process.\\
\textbf{\code{pops}} & A character vector giving the population variable. Only used if \code{pop_adjust = TRUE}.\\
\textbf{\code{prior\_susc}} & Prior on $S_{v-1} / P$, the initial susceptible population as a proportion of the population size. If \code{NULL}, this is assumed to be equal to 1 (i.e. everyone is initially susceptible). Otherwise, can be a call to \code{normal()} from \pkg{rstanarm}, which assigns a normal prior truncated to $[0,1]$. Only used if \code{pop_adjust = TRUE}.\\
\addlinespace
\textbf{\code{rm}} & A character vector giving the variable corresponding to $v_t$, i.e. the proportion of $S_t$ to remove at time $t$. Only used if \code{pop_adjust = TRUE}.\\
\textbf{\code{prior\_rm\_noise}} & Prior on the parameter $\xi$, which controls noise around $v_t$. If \code{NULL}, no noise is added. Only used if \code{pop_adjust = TRUE}.\\
\bottomrule
\end{tabular}
\end{table}

\end{CodeChunk}

\hypertarget{sec:imp_observations}{%
\subsection{Observations}\label{sec:imp_observations}}

An observational model is defined by a call to \code{epiobs()}. In
particular, this must also make explicit the model for the multipliers
\(\alpha_t\), and must also specify the coefficients \(\pi_k\).
\code{epiobs()} has a \code{formula} argument. The left hand side must
indicate the observation vector to be modeled, while the right hand side
defines a linear predictor for \(\alpha_t\). The argument \code{i2o}
plays a similar role to the \code{gen} argument in \code{epiinf()},
however it instead corresponds the vector \(\pi\) in Equation
\eqref{eq:obsmean}.

Take for example the task of modeling daily \code{deaths}, which as we
saw is a variable in \code{data}. A possible model is the following.

\begin{CodeChunk}
\begin{CodeInput}
R> deaths <- epiobs(formula = deaths ~ 1, i2o = EuropeCovid$inf2death, 
+                  link = scaled_logit(0.02))
\end{CodeInput}
\end{CodeChunk}

Here \(\alpha_t\) corresponds to the infection fatality rate (IFR), and
is modeled as an intercept transformed by the scaled-logit link. This
implies that the IFR is constant over time and its value lies somewhere
between 0\% and 2\%. If the prior on the intercept (specified by the
\code{prior_intercept} argument) is chosen to be symmetric around zero,
then the prior mean for the IFR is 1\%. \code{EuropeCovid$inf2death} is
a numeric simplex vector that gives the same delay distribution as used
in \citet{Flaxman2020}. This is a density function for a discretized
mixture of Gamma random variables.

Additional arguments include \code{family}, which specifies the sampling
distribution \(p(y_t, \phi)\). There are also arguments allowing the
user to control prior distributions for effects in the linear predictor
for \(\alpha_t\), and the prior on the auxiliary variable \(\phi\). All
arguments to \code{epiobs()} are shown in Table
\ref{tab:model-imp-epiobs-args}.

\begin{CodeChunk}
\begin{table}[!h]

\caption{\label{tab:model-imp-epiobs-args}\small Formal arguments for \code{epiobs()}. This defines a single observation model. Multiple such models can be used and passed to \code{epim()} in a list.}
\centering
\begin{tabular}[t]{>{\raggedright\arraybackslash}p{7em}>{\raggedright\arraybackslash}p{30em}}
\toprule
Argument & Description\\
\midrule
\code{formula} & An object of class \code{"formula"} which determines the linear predictor for the ascertainment rate. The left hand side must define the response that is being modeled (i.e. the actual observations, not the latent ascertainments)\\
\code{i2o} & A numeric (probability) vector defining the probability mass function $\pi_k$ of the time from an infection to an observation.\\
\code{family} & A string representing the family of the sampling distribution $p(y_t,\phi)$. Can be one of \code{"poisson"}, \code{"neg_binom"}, \code{"quasi_poisson"}, \code{"normal"} or \code{"log_normal"}.\\
\code{link} & A string representing the link function used to transform the linear predictor. Can be one of \code{"logit"}, \code{"probit"}, \code{"cauchit"}, \code{"cloglog"}, \code{"identity"}. Defaults to \code{"logit"}.\\
\code{center},
 \code{prior},
 \code{prior\_intercept} & same as in \code{epirt()}, described above.\\
\addlinespace
\code{prior\_aux} & The prior distribution for the auxiliary parameter $\phi$, if it exists. Only used if family is \code{"neg_binom"} (reciprocal dispersion), \code{"quasi_poisson"} (dispersion), \code{"normal"} (standard deviation) or \code{"log_normal"} (sigma parameter).\\
\code{...} & Additional arguments for \code{model.frame()} from \pkg{stats}.\\
\bottomrule
\end{tabular}
\end{table}

\end{CodeChunk}

\hypertarget{sec:data}{%
\subsection{Data}\label{sec:data}}

Before fitting our first model in Section \ref{sec:first_fit}, we
elaborate on the \code{data} argument to \code{epim()}. Recall that this
must contain all variables used in the transmission and infection
models, and in all observational models. For our example, \code{data}
looks like

\begin{CodeChunk}
\begin{CodeInput}
R> head(data)
\end{CodeInput}
\begin{CodeOutput}
# A tibble: 6 x 9
# Groups:   country [1]
  country date       schools_universiti~ self_isolating_if_~
  <fct>   <date>                   <int>               <int>
1 Austria 2020-02-22                   0                   0
2 Austria 2020-02-23                   0                   0
3 Austria 2020-02-24                   0                   0
4 Austria 2020-02-25                   0                   0
5 Austria 2020-02-26                   0                   0
6 Austria 2020-02-27                   0                   0
# ... with 5 more variables: public_events <int>,
#   lockdown <int>, social_distancing_encouraged <int>,
#   deaths <int>, pop <int>
\end{CodeOutput}
\end{CodeChunk}

The columns \code{country} and \code{date} define the region and time
period corresponding to each of the remaining variables. \code{epim()}
assumes that the first seeding day (i.e.~the start of the epidemic) in
each region is the first date found in the data frame. The last data
found for each region is the final data at which the epidemic is
simulated. It is up to the user to appropriately choose these dates. For
our example, the first and last dates for each group can be seen as
follows.

\begin{CodeChunk}
\begin{CodeInput}
R> dates <- summarise(data, start = min(date), end = max(date))
R> head(dates)
\end{CodeInput}
\begin{CodeOutput}
# A tibble: 6 x 3
  country start      end       
  <fct>   <date>     <date>    
1 Austria 2020-02-22 2020-05-05
2 Belgium 2020-02-18 2020-05-05
3 Denmark 2020-02-21 2020-05-05
4 France  2020-02-07 2020-05-05
5 Germany 2020-02-15 2020-05-05
6 Italy   2020-01-27 2020-05-05
\end{CodeOutput}
\end{CodeChunk}

Here, the start dates have been heuristically chosen to be 30 days prior
to observing 10 cumulative deaths in each country.

\hypertarget{sec:first_fit}{%
\subsection{A First Fit}\label{sec:first_fit}}

We are now ready to fit our first model. For this we return to the model
fitting function \code{epim()}. The following command is used to
instruct \pkg{epidemia} to run Markov chains in parallel, rather than
sequentially, if multiple cores are detected.

\begin{CodeChunk}
\begin{CodeInput}
R>  options(mc.cores = parallel::detectCores())
\end{CodeInput}
\end{CodeChunk}

Our call to \code{epim()} is as follows. We use \code{refresh = 0} to
suppress printing output in this article,however, this should not
generally be used as such output is useful.

\begin{CodeChunk}
\begin{CodeInput}
R> fm <- epim(rt = rt, inf = inf, obs = deaths, data = data, 
+            group_subset = "France", algorithm = "sampling", iter = 1e3, 
+            seed = 12345, refresh = 0)
\end{CodeInput}
\end{CodeChunk}

The print method for \code{epimodel} objects prints summary statistics
for model parameters. These are obtained from the sampled posterior
distribution. Parameter are displayed according to which part of the
model they belong to (transmission, observations, infections). An
estimate of the standard deviation, labeled \code{MAD_SD} is displayed.
This is the median absolute deviation from the median, and is more
robust than naive estimates of the standard deviation for long-tailed
distributions.

\begin{CodeChunk}
\begin{CodeInput}
R> print(fm)
\end{CodeInput}
\begin{CodeOutput}

Rt regression parameters:
==========
coefficients:
                Median MAD_SD
R|(Intercept)    0.7    0.2  
R|lockdown      -2.4    0.3  
R|public_events -0.4    0.3  

 deaths  regression parameters:
==========
coefficients:
                             Median MAD_SD
deaths|(Intercept)            0.0    0.2  
deaths|reciprocal dispersion 10.4    0.4  

Infection model parameters:
==========
              Median MAD_SD
seeds[France] 15.2    5.2  
seeds_aux     27.3   22.0  
\end{CodeOutput}
\end{CodeChunk}

Alternatively, the summary method can be used. This gives quantiles of
the posterior draws, and also displays some MCMC diagnostics.

\begin{CodeChunk}
\begin{CodeInput}
R> summary(fm)
\end{CodeInput}
\begin{CodeOutput}


Estimates:
                               mean   sd   10%   50%   90%
R|(Intercept)                 0.7    0.2  0.5   0.7   0.9 
R|lockdown                   -2.4    0.3 -2.8  -2.4  -2.1 
R|public_events              -0.4    0.3 -0.8  -0.4   0.0 
deaths|(Intercept)            0.0    0.2 -0.3   0.0   0.2 
seeds[France]                16.2    5.9  9.6  15.2  24.0 
seeds_aux                    39.7   38.0  8.7  27.3  84.8 
deaths|reciprocal dispersion 10.5    0.5 10.1  10.4  11.1 

MCMC diagnostics
                             mcse Rhat n_eff
R|(Intercept)                0.0  1.0  1110 
R|lockdown                   0.0  1.0  1119 
R|public_events              0.0  1.0   921 
deaths|(Intercept)           0.0  1.0  1422 
seeds[France]                0.2  1.0  1297 
seeds_aux                    1.2  1.0  1061 
deaths|reciprocal dispersion 0.0  1.0  2238 
log-posterior                0.1  1.0   738 
\end{CodeOutput}
\end{CodeChunk}

\hypertarget{sec:epi_examples}{%
\section{Examples}\label{sec:epi_examples}}

\hypertarget{sec:flu}{%
\subsection{Spanish Flu in Baltimore}\label{sec:flu}}

Our first example infers \(R_t\) during the H1N1 pandemic in Baltimore
in 1918, using only case counts and a serial interval. This is,
relatively speaking, a simple setting for several reasons. Only a single
population (that of Baltimore) and observational model (case data) are
considered. \(R_t\) will follow a daily random walk with no additional
covariates. Of course, \pkg{epidemia} is capable of more complex
modeling, and Section \ref{sec:epi_multilevel} takes a step in this
direction.

In addition to inferring \(R_t\), this example demonstrates how to
undertake posterior predictive checks to graphically assess model fit.
The basic model outlined above is then extended to add variation to the
infection process, as was outlined in Section \ref{sec:latent}. This is
particularly useful for this example because infection counts are low.
We will also see that the extended model appears to have a computational
advantage in this setting.

The case data is provided by the \proglang{R} package \pkg{EpiEstim}.

\begin{CodeChunk}
\begin{CodeInput}
R> library(EpiEstim)
R> data("Flu1918")
R> print(Flu1918)
\end{CodeInput}
\begin{CodeOutput}
$incidence
 [1]   5   1   6  15   2   3   8   7   2  15   4  17   4  10
[15]  31  11  13  36  13  33  17  15  32  27  70  58  32  69
[29]  54  80 405 192 243 204 280 229 304 265 196 372 158 222
[43] 141 172 553 148  95 144  85 143  87  73  70  62 116  44
[57]  38  60  45  60  27  51  34  22  16  11  18  11  10   8
[71]  13   3   3   6   6  13   5   6   6   5   5   1   2   2
[85]   3   8   4   1   2   3   1   0

$si_distr
 [1] 0.000 0.233 0.359 0.198 0.103 0.053 0.027 0.014 0.007
[10] 0.003 0.002 0.001
\end{CodeOutput}
\end{CodeChunk}

\hypertarget{data}{%
\subsubsection{Data}\label{data}}

First form the \code{data} argument, which will eventually be passed to
the model fitting function \code{epim()}. Recall that this must be a
data frame containing all observations and covariates used to fit the
model. Therefore, we require a column giving cases over time. In this
example, no covariates are required. \(R_t\) follows a daily random
walk, with no additional covariates. In addition, the case ascertainment
rate will be assumed at 100\%, and so no covariates are used for this
model either.

\begin{CodeChunk}
\begin{CodeInput}
R> date <- as.Date("1918-01-01") + seq(0, along.with = c(NA, Flu1918$incidence))
R> data <- data.frame(city = "Baltimore", cases = c(NA, Flu1918$incidence), 
+                    date = date)
\end{CodeInput}
\end{CodeChunk}

The variable \code{date} has been constructed so that the first cases
are seen on the second day of the epidemic rather than the first. This
ensures that the first observation can be explained by past infections.

\hypertarget{transmission}{%
\subsubsection{Transmission}\label{transmission}}

Recall that we wish to model \(R_t\) by a daily random walk. This is
specified by a call to \code{epirt()}. The \code{formula} argument
defines the linear predictor which is then transformed by the link
function. A random walk can be added to the predictor using the
\code{rw()} function. This has an optional \code{time} argument which
allows the random walk increments to occur at a different frequency to
the \code{date} column. This can be employed, for example, to define a
weekly random walk. If unspecified, the increments are daily. The
increments are modeled as half-normal with a scale hyperparameter. The
value of this is set using the \code{prior_scale} argument. This is used
in the snippet below.

\begin{CodeChunk}
\begin{CodeInput}
R> rt <- epirt(formula = R(city, date) ~ 1 + rw(prior_scale = 0.01),
+             prior_intercept = normal(log(2), 0.2), link = 'log')
\end{CodeInput}
\end{CodeChunk}

The prior on the intercept gives the initial reproduction number \(R_0\)
a prior mean of roughly 2.

\hypertarget{observations}{%
\subsubsection{Observations}\label{observations}}

Multiple observational models can be collected into a list and passed to
\code{epim()} as the \code{obs} argument. In this case, only case data
is used and so there is only one such model.

\begin{CodeChunk}
\begin{CodeInput}
R> obs <-  epiobs(formula = cases ~ 0 + offset(rep(1,93)), link = "identity", 
+                i2o = rep(.25,4))
\end{CodeInput}
\end{CodeChunk}

For the purpose of this exercise, we have assumed that all infections
will eventually manifest as a case. The above snippet implies
\textit{full} ascertainment, i.e.~\(\alpha_t = 1\) for all \(t\). This
is achieved using \code{offset()}, which allows vectors to be added to
the linear predictor without multiplication by an unknown parameter.

The \code{i2o} argument implies that cases are recorded with equal
probability in any of the four days after infection.

\hypertarget{infections}{%
\subsubsection{Infections}\label{infections}}

Two infection models are considered. The first uses the renewal equation
(Equation \ref{eq:renewal2}) to propagate infections. The extended model
adds variance to this process, and can be applied by using
\code{latent = TRUE} in the call to \code{epiinf()}.

\begin{CodeChunk}
\begin{CodeInput}
R> inf <- epiinf(gen = Flu1918$si_distr)
R> inf_ext <-  epiinf(gen = Flu1918$si_distr, latent = TRUE, 
+                    prior_aux = normal(10,2))
\end{CodeInput}
\end{CodeChunk}

The argument \code{gen} takes a discrete generation distribution. Here
we have used the serial interval provided by \pkg{EpiEstim}. As in
Section \ref{sec:imp_infections}, this makes the implicit assumption
that the serial interval approximates the generation time.
\code{prior_aux} sets the prior on the coefficient of dispersion \(d\).
This prior assumes that infections have conditional variance around 10
times the conditional mean.

\hypertarget{fitting-the-model}{%
\subsubsection{Fitting the Model}\label{fitting-the-model}}

We are left to collect all remaining arguments required for
\code{epim()}. This is done as follows.

\begin{CodeChunk}
\begin{CodeInput}
R> args <- list(rt = rt, obs = obs, inf = inf, data = data, iter = 2e3, 
+              seed = 12345)
R> args_ext <- args; args_ext$inf <- inf_ext
\end{CodeInput}
\end{CodeChunk}

The arguments \code{iter} and \code{seed} set the number of MCMC
iterations and seeds respectively, and are passed directly on to the
\code{sampling()} function from \pkg{rstan}.

We wrap the calls to \code{epim()} in \code{system.time} in order to
assess the computational cost of fitting the models. The snippet below
fits both versions of the model. \code{fm1} and \code{fm2} are the
fitted basic model and extended model respectively.

\begin{CodeChunk}
\begin{CodeInput}
R> system.time(fm1 <- do.call(epim, args))
\end{CodeInput}
\begin{CodeOutput}
   user  system elapsed 
341.488   1.550  93.214 
\end{CodeOutput}
\begin{CodeInput}
R> system.time(fm2 <- do.call(epim, args_ext))
\end{CodeInput}
\begin{CodeOutput}
   user  system elapsed 
 55.377   0.895  17.583 
\end{CodeOutput}
\end{CodeChunk}

Note the stark difference in running time. The extended model appears to
fit faster even though there are \(87\) additional parameters being
sampled (daily infections after the seeding period, and the coefficient
of dispersion). We conjecture that the additional variance around
infections adds slack to the model, and leads to a posterior
distribution that is easier to sample.

\begin{figure}[H]
    \centering
    \begin{adjustbox}{width= 0.90\textwidth}
    \includegraphics{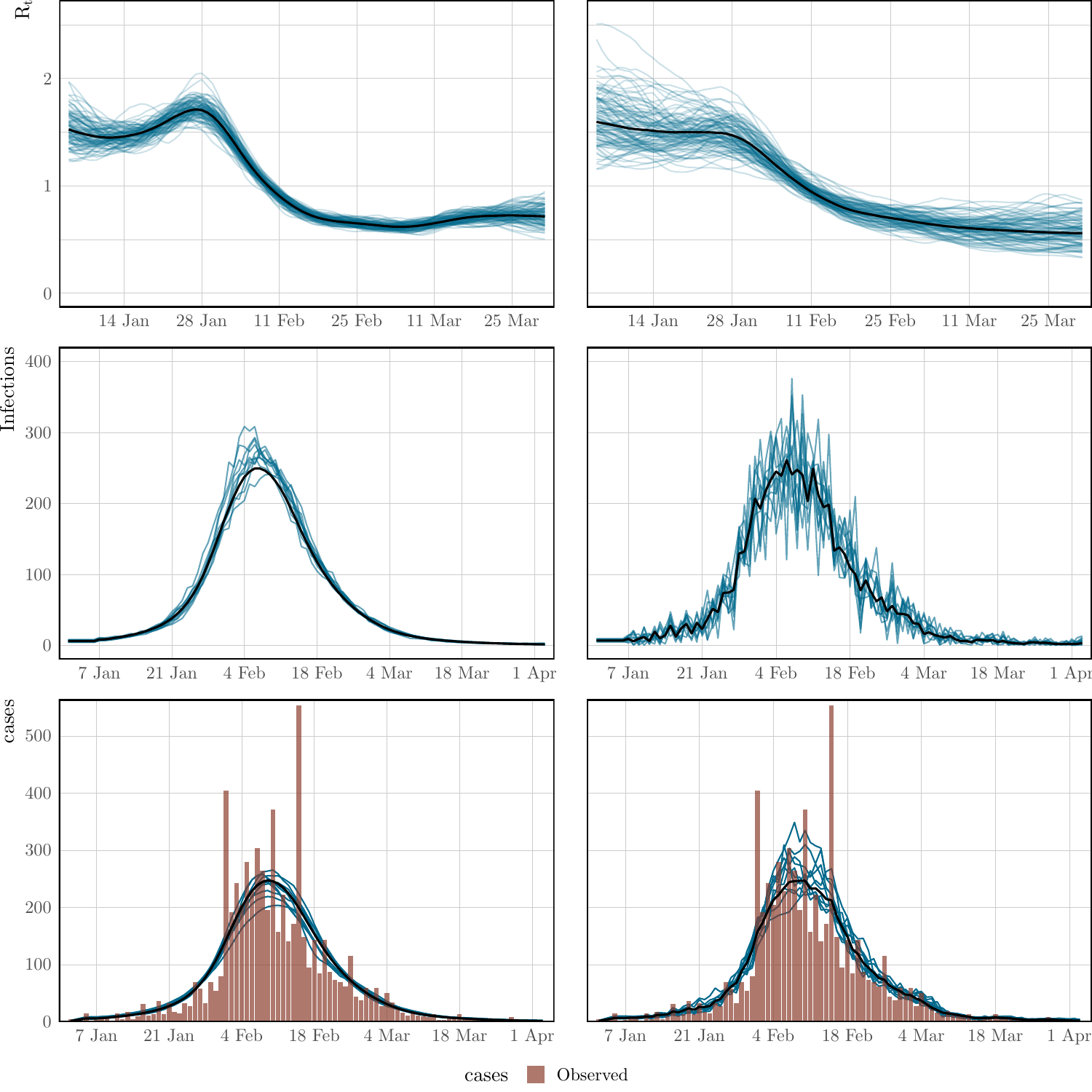}
    \end{adjustbox}
    \caption{\small Spaghetti plots showing the median (black) and sample paths (blue) from the posterior distribution. These result from the model fits in Section \ref{sec:flu}. Left corresponds to the basic model, and the right panel is for the extended version. Top: Inferred time-varying reproduction numbers, which have been smoothed over 7 days for illustration. Middle: Inferred latent infections. Bottom: Observed cases, and cases simulated from the posterior. These align closely, and so do not flag problems with the model fit.}
    \label{fig:flu1918-plots}
\end{figure}

The results of both models are shown in Figure \ref{fig:flu1918-plots}.
This Figure has been produced using \pkg{epidemia}'s plotting functions.
The key difference stems from the infection process. In the basic model,
infections are deterministic given \(R_t\) and seeds. However when
infection counts are low, we generally expect high variance in the
infection process. Since this variance is unaccounted for, the model
appears to place too much confidence in \(R_t\) in this setting. The
extended model on the other hand, has much wider credible intervals for
\(R_t\) when infections are low. This is intuitive: when counts are low,
changes in infections could be explained by either the variance in the
offspring distribution of those who are infected, or by changes in the
\(R_t\) value. This model captures this intuition.

Posterior predictive checks are shown in the bottom panel of Figure
\ref{fig:flu1918-plots}, and show that both models are able to fit the
data well.

\hypertarget{sec:epi_multilevel}{%
\subsection{Assessing the Effects of Interventions on
COVID-19}\label{sec:epi_multilevel}}

The Spanish flu example (Section \ref{sec:flu}) considered inferring the
instantaneous reproduction number over time in a single population.
Here, we demonstrate some of the more advanced modeling capabilities of
the package.

Consider modeling the evolution of an epidemic in multiple distinct
regions. As discussed in Section \ref{sec:mult}, one can always approach
this by modeling each group separately. It was argued that this approach
is fast, because models may be fit independently. Nonetheless, often
there is little high quality data for some groups, and the data does
little to inform parameter estimates. This is particularly true in the
early stages of an epidemic. Joining regions together through
hierarchical models allows information to be shared between regions in a
natural way, improving parameter estimates while still permitting
between group variation.

In this section, we use a hierarchical model to estimate the effect of
non-pharmaceutical interventions (NPIs) on the transmissibility of
Covid-19. We consider the same setup as \citet{Flaxman2020}: attempting
to estimate the effect of a set of measures that were implemented in
March 2020 in 11 European countries during the first wave of Covid-19.
This will be done by fitting the model to daily death data. The same set
of measures and countries that were used in \citet{Flaxman2020} are also
used here. \citet{Flaxman2020MA} considered a version of this model that
used partial pooling for all NPI effects. Here, we consider a model that
uses the same approach.

This example is not intended to be a fully rigorous statistical
analysis. Rather, the intention is to demonstrate partial pooling of
parameters in \pkg{epidemia} and how to infer their effect sizes. We
also show how to forecast observations into the future, and how to
undertake counterfactual analyses.

\hypertarget{data-1}{%
\subsubsection{Data}\label{data-1}}

We use a data set \code{EuropeCovid2}, which is provided by
\pkg{epidemia}. This contains daily death and case data in the 11
countries concerned up until the 1\textsuperscript{st} July 2020. The
data derives from the WHO COVID-19 explorer as of the
5\textsuperscript{th} of January 2021. This differs from the data used
in \citet{Flaxman2020}, because case and death counts have been adjusted
retrospectively as new information came to light. \pkg{epidemia} also
has a data set \code{EuropeCovid} which contains the same data as that
in \citet{Flaxman2020}, and this could alternatively be used for this
exercise.

\code{EuropeCovid2} also contains binary series representing the set of
five mitigation measures considered in \citet{Flaxman2020}. These
correspond to the closing of schools and universities, the banning of
public events, encouraging social distancing, requiring self isolation
if ill, and finally the implementation of full lockdown. The dates at
which these policies were enacted are exactly the same as those used in
\citet{Flaxman2020}.

Load the data set as follows.

\begin{CodeChunk}
\begin{CodeInput}
R> data("EuropeCovid2")
R> data <- EuropeCovid2$data
R> head(data)
\end{CodeInput}
\begin{CodeOutput}
# A tibble: 6 x 11
# Groups:   country [1]
  id    country date       cases deaths schools_universities
  <chr> <chr>   <date>     <int>  <int>                <int>
1 AT    Austria 2020-01-03     0      0                    0
2 AT    Austria 2020-01-04     0      0                    0
3 AT    Austria 2020-01-05     0      0                    0
4 AT    Austria 2020-01-06     0      0                    0
5 AT    Austria 2020-01-07     0      0                    0
6 AT    Austria 2020-01-08     0      0                    0
# ... with 5 more variables: self_isolating_if_ill <int>,
#   public_events <int>, lockdown <int>,
#   social_distancing_encouraged <int>, pop <int>
\end{CodeOutput}
\end{CodeChunk}

Recall that for each country, \pkg{epidemia} will use the earliest date
in \code{data} as the first date to begin seeding infections. Therefore,
we must choose an appropriate start date for each group. One option is
to use the same rule as in \citet{Flaxman2020}, and assume that seeding
begins in each country 30 days prior to observing 10 cumulative deaths.
To do this, we filter the data frame as follows.

\begin{CodeChunk}
\begin{CodeInput}
R> data <- filter(data, date > date[which(cumsum(deaths) > 10)[1] - 30])
\end{CodeInput}
\end{CodeChunk}

This leaves the following assumed start dates.

\begin{CodeChunk}
\begin{CodeInput}
R> dates <- summarise(data, start = min(date), end = max(date))
R> head(dates)
\end{CodeInput}
\begin{CodeOutput}
# A tibble: 6 x 3
  country start      end       
  <chr>   <date>     <date>    
1 Austria 2020-02-23 2020-06-30
2 Belgium 2020-02-15 2020-06-30
3 Denmark 2020-02-22 2020-06-30
4 France  2020-02-09 2020-06-30
5 Germany 2020-02-16 2020-06-30
6 Italy   2020-01-28 2020-06-30
\end{CodeOutput}
\end{CodeChunk}

Although \code{data} contains observations up until the end of June, we
fit the model using a subset of the data. We hold out the rest to
demonstrate forecasting out-of-sample. Following \citet{Flaxman2020},
the final date considered is the 5\textsuperscript{th} May.

\begin{CodeChunk}
\begin{CodeInput}
R> data <- filter(data, date < as.Date("2020-05-05"))
\end{CodeInput}
\end{CodeChunk}

\hypertarget{model-components}{%
\subsubsection{Model Components}\label{model-components}}

We have seen several times now that \pkg{epidemia} require the user to
specify three model components: transmission, infections, and
observations. These are now considered in turn.

\hypertarget{sec:europe-covid_transmission}{%
\paragraph{Transmission}\label{sec:europe-covid_transmission}}

Country-specific reproduction numbers \(R^{(m)}_{t}\) are expressed in
terms of the control measures. Since the measures are encoded as binary
policy indicators, reproduction rates must follow a step function. They
are constant between policies, and either increase or decrease as
policies come into play. The implicit assumption, of course, is that
only control measures may affect transmission, and that these effects
are fully realized instantaneously.

Let \(t^{(m)}_{k} \geq 0\), \(k \in \{1,\ldots,5\}\) be the set of
integer times at which the \(k\)\textsuperscript{th} control measure was
enacted in the \(m\)\textsuperscript{th} country. Accordingly, we let
\(I^{(m)}_{k}\), \(k \in \{1,\ldots,5\}\) be a set of corresponding
binary vectors such that \begin{equation}
I^{(m)}_{k,t} = 
\begin{cases}
0, & \text{if } t < t^{(m)}_k \\
1. & \text{if } t \geq t^{(m)}_k
\end{cases}
\end{equation} Reproduction numbers are mathematically expressed as
\begin{equation}
R^{(m)}_{t} = R' g^{-1}\left(b^{(m)}_0 + \sum_{k=1}^{5}\left(\beta_k + b^{(m)}_k\right)I^{(m)}_{k,t}\right),
\end{equation} where \(R' = 3.25\) and \(g\) is the logit-link.
Parameters \(b^{(m)}_0\) are country-specific intercepts, and each
\(b^{(m)}_k\) is a country effect for the \(k\)\textsuperscript{th}
measure. The intercepts allow each country to have its own initial
reproduction number, and hence accounts for possible variation in the
inherent transmissibility of Covid-19 in each population. \(\beta_k\) is
a fixed effect for the \(k\)\textsuperscript{th} policy. This quantity
corresponds to the average effect of a measure across all countries
considered.

Control measures were implemented in quick succession in most countries.
For some countries, a subset of the measures were in fact enacted
simultaneously. For example, Germany banned public events at the same
time as implementing lockdown. The upshot of this is that policy effects
are \emph{highly colinear} and may prove difficult to infer with
uninformative priors.

One potential remedy is to use domain knowledge to incorporate
information into the priors. In particular, it seems a priori unlikely
that the measures served to increase transmission rates significantly.
It is plausible, however, that each had a significant effect on reducing
transmission. A symmetric prior like the Gaussian does not capture this
intuition and increases the difficulty in inferring effects, because
they are more able to offset each other. This motivated the prior used
in \citet{Flaxman2020}, which was a Gamma distribution shifted to have
support other than zero.

We use the same prior in our example. Denoting the distribution of a
Gamma random variable with shape \(a\) and scale \(b\) by
\(\text{Gamma}(a, b)\), this prior is \begin{equation}
-\beta_k - \frac{\log(1.05)}{6} \sim \text{Gamma}(1/6, 1).
\label{eq:betaprior}
\end{equation} The shift allows the measures to increase transmission
slightly.

All country-specific parameters are partially pooled by letting
\begin{equation}
b^{(m)}_k \sim N(0, \sigma_k),
\end{equation} where \(\sigma_k\) are standard deviations,
\(\sigma_0 \sim \text{Gamma}(2, 0.25)\) and
\(\sigma_k \sim \text{Gamma}(0.5, 0.25)\) for all \(k > 0\). This gives
the intercept terms more variability under the prior.

The transmission model described above is expressed programmatically as
follows.

\begin{CodeChunk}
\begin{CodeInput}
R> rt <- epirt(formula = R(country, date) ~ 0 + (1 + public_events + 
+               schools_universities + self_isolating_if_ill + 
+               social_distancing_encouraged + lockdown || country) + 
+               public_events + schools_universities + self_isolating_if_ill +
+               social_distancing_encouraged + lockdown, 
+             prior = shifted_gamma(shape = 1/6, scale = 1, shift = log(1.05)/6),
+             prior_covariance = decov(shape = c(2, rep(0.5, 5)), scale = 0.25),
+             link = scaled_logit(6.5))
\end{CodeInput}
\end{CodeChunk}

The operator \code{||} is used rather than \code{|} for random effects.
This ensures that all effects for a given country are independent, as
was assumed in the model described above. Using \code{|} would
alternatively give a prior on the full covariance matrix, rather than on
the individual \(\sigma_i\) terms. The argument \code{prior} reflects
Equation \eqref{eq:betaprior}. Since country effects are assumed
independent, the \code{decov} prior reduces to assigning Gamma priors to
each \(\sigma_i\). By using a vector rather than a scalar for the
\code{shape} argument, we are able to give the prior on the intercepts a
larger shape parameter.

\hypertarget{infections-1}{%
\paragraph{Infections}\label{infections-1}}

Infections are kept simple here by using the basic version of the model.
That is to say that infections are taken to be a deterministic function
of seeds and reproduction numbers, propagated by the renewal process.
Extensions to modeling infections as parameters and adjustments for the
susceptible population are not considered. The model is defined as
follows.

\begin{CodeChunk}
\begin{CodeInput}
R> inf <- epiinf(gen = EuropeCovid$si, seed_days = 6)
\end{CodeInput}
\end{CodeChunk}

\code{EuropeCovid$si} is a numeric vector giving the serial interval
used in \citet{Flaxman2020}. As in that work, we make no distinction
between the generation distribution and serial interval here.

\hypertarget{observations-1}{%
\paragraph{Observations}\label{observations-1}}

In order to infer the effects of control measures on transmission, we
must fit the model to data. Here, daily deaths are used. In theory,
additional types of data can be included in the model, but such
extension are not considered here. A simple intercept model is used for
the infection fatality rate (IFR). This makes the assumption that the
IFR is constant over time. The model can be written as follows.

\begin{CodeChunk}
\begin{CodeInput}
R> deaths <- epiobs(formula = deaths ~ 1, i2o = EuropeCovid2$inf2death, 
+                  prior_intercept = normal(0,0.2), link = scaled_logit(0.02))
\end{CodeInput}
\end{CodeChunk}

By using \code{link = scaled_logit(0.02)}, we let the IFR range between
\(0\%\) and \(2\%\). In conjunction with the symmetric prior on the
intercept, this gives the IFR a prior mean of \(1\%\).
\code{EuropeCovid2$inf2death} is a numeric vector giving the same
distribution for the time from infection to death as that used in
\citet{Flaxman2020}.

\hypertarget{model-fitting}{%
\subsubsection{Model Fitting}\label{model-fitting}}

In general, \pkg{epidemia}'s models should be fit using Hamiltonian
Monte Carlo. For this example, however, we use Variational Bayes (VB) as
opposed to full MCMC sampling. This is because full MCMC sampling of a
joint model of this size is computationally demanding, due in part to
renewal equation having to be evaluated for each region and for each
evaluation of the likelihood and its derivatives. Nonetheless, VB allows
rapid iteration of models and may lead to reasonable estimates of effect
sizes. For this example, we have also run full MCMC, and the inferences
reported here are not substantially different.

\hypertarget{prior-check}{%
\paragraph{Prior Check}\label{prior-check}}

Section \ref{sec:flu} gave an example of using posterior predictive
checks. It is also useful to do prior predictive checks as these allow
the user to catch obvious mistakes that can occur when specifying the
model, and can also help to affirm that the prior is in fact reasonable.

In \pkg{epidemia} we can do this by using the \code{priorPD = TRUE} flag
in \code{epim()}. This discards the likelihood component of the
posterior, leaving just the prior. We use Hamiltonian Monte Carlo over
VB for the prior check, partly because sampling from the prior is quick
(it is the likelihood that is expensive to evaluate). In addition, we
have defined Gamma priors on some coefficients, which are generally
poorly approximated by VB.

\begin{CodeChunk}
\begin{CodeInput}
R> args <- list(rt = rt, inf = inf, obs = deaths, data = data, seed = 12345, 
+              refresh = 0)
R> pr_args <- c(args, list(algorithm = "sampling", iter = 1e3, prior_PD = TRUE))
R> fm_prior <- do.call(epim, pr_args)
\end{CodeInput}
\end{CodeChunk}

\begin{figure}[H]
    \centering
    \begin{adjustbox}{width= 0.8\textwidth}
    \includegraphics{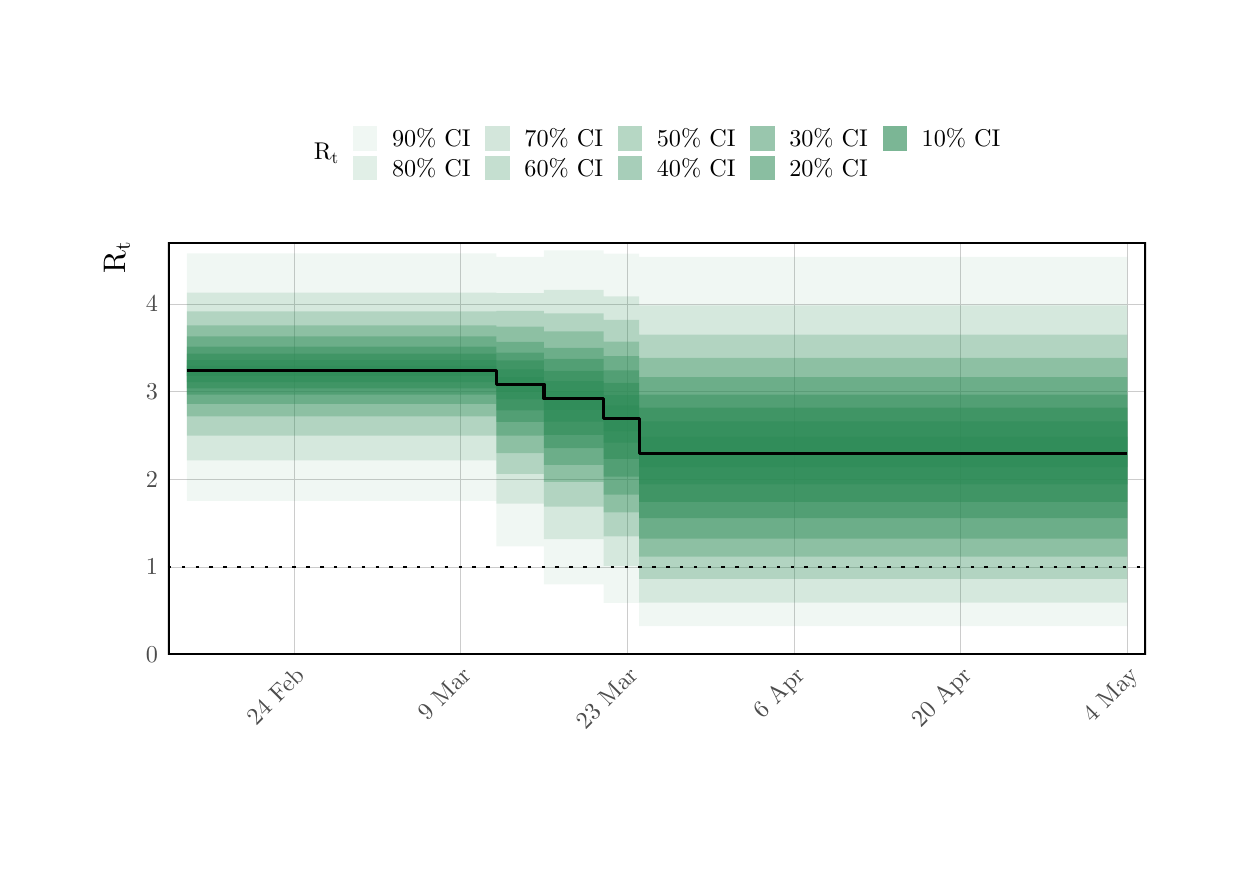}
    \end{adjustbox}
    \vspace{-10mm}
    \caption{\small A prior predictive check for reproduction numbers $R_t$ in the multilevel model of Section \ref{sec:epi_multilevel}. Only results for the United Kingdom are presented. The prior median is shown in black, with credible intervals shown in various shades of green. The check appears to confirm that $R_t$ follows a step-function, as we expect given the definition in Section \ref{sec:europe-covid_transmission}.}
    \label{fig:multilevel-prior}
\end{figure}

Figure \ref{fig:multilevel-prior} shows approximate samples of
\(R_{t,m}\) from the prior distribution. This confirms that reproduction
numbers follow a step function, and that rates can both increase and
decrease as measures come into play.

\hypertarget{approximating-the-posterior}{%
\paragraph{Approximating the
Posterior}\label{approximating-the-posterior}}

The model will be fit using Variational Bayes by using
\code{algorithm = "fullrank"} in the call to \code{epim()}. This is
generally preferable to \code{"meanfield"} for these models, largely
because \code{"meanfield"} ignores posterior correlations. We decrease
the parameter \code{tol_rel_obj} from its default value, and increase
the number of iterations to aid convergence.

\begin{CodeChunk}
\begin{CodeInput}
R> args$algorithm <- "fullrank"; args$iter <- 5e4; args$tol_rel_obj <- 1e-3
R> fm <- do.call(epim, args)
\end{CodeInput}
\end{CodeChunk}

\begin{figure}[ht]
    \centering
    \begin{adjustbox}{width=0.9\textwidth}
     \includegraphics{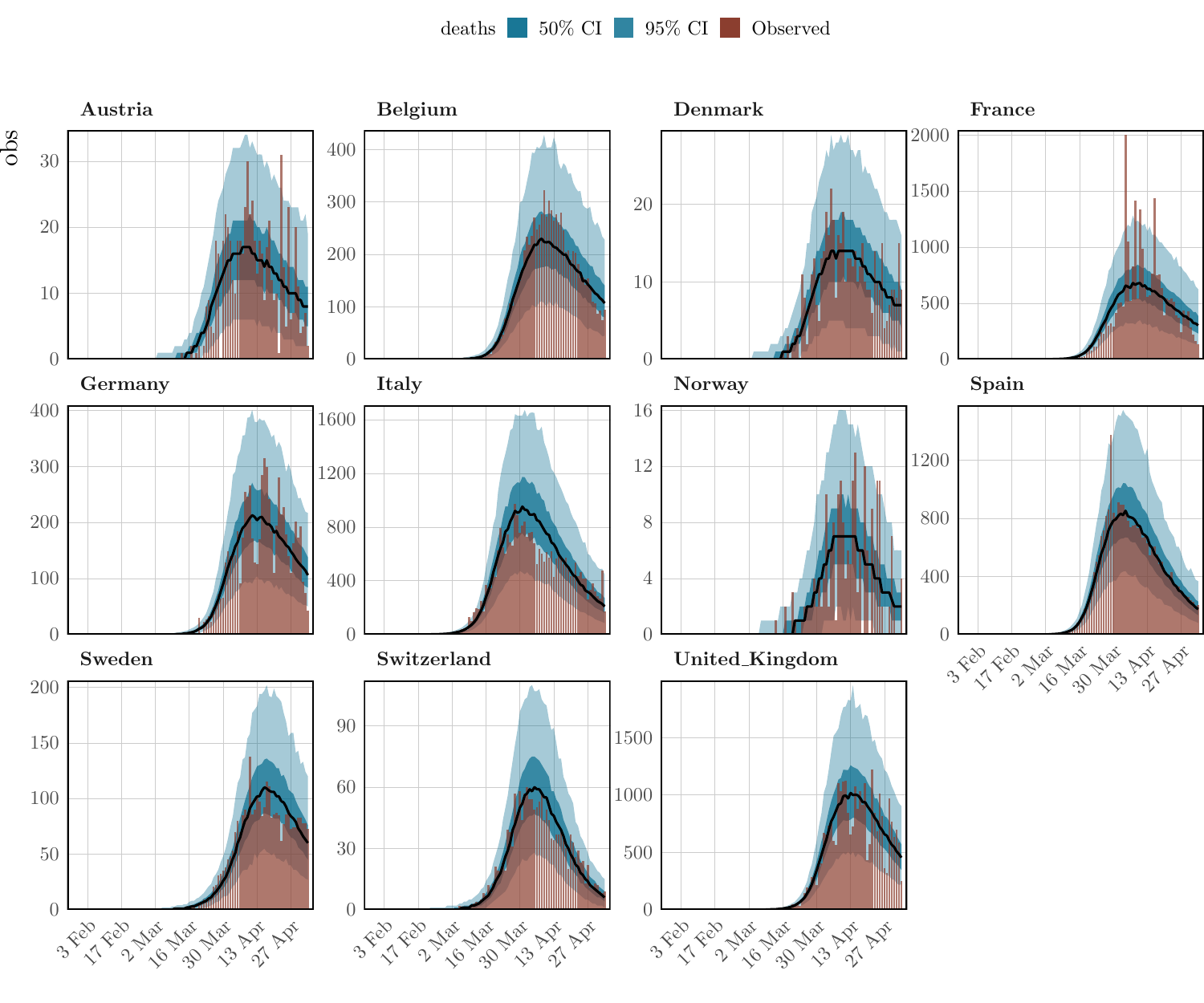}
    \end{adjustbox}
    \caption{\small Posterior predictive checks for the multilevel model. Observed daily deaths (red) is plotted as a bar plot. Credible intervals from the posterior are plotted in shades of blue, in addition to the posterior median in black.}
    \label{fig:multilevel-obs-plots}
\end{figure}

A first step in evaluating the model fit is to perform posterior
predictive checks. This is to confirm that the model adequately explains
the observed daily deaths in each region. This can be done using the
command \code{plot_obs(fm, type = "deaths", levels = c(50, 95))}. The
plot is shown in Figure \ref{fig:multilevel-obs-plots}.

\begin{figure}[ht]
    \centering
    \begin{adjustbox}{width=0.9\textwidth}
    \includegraphics{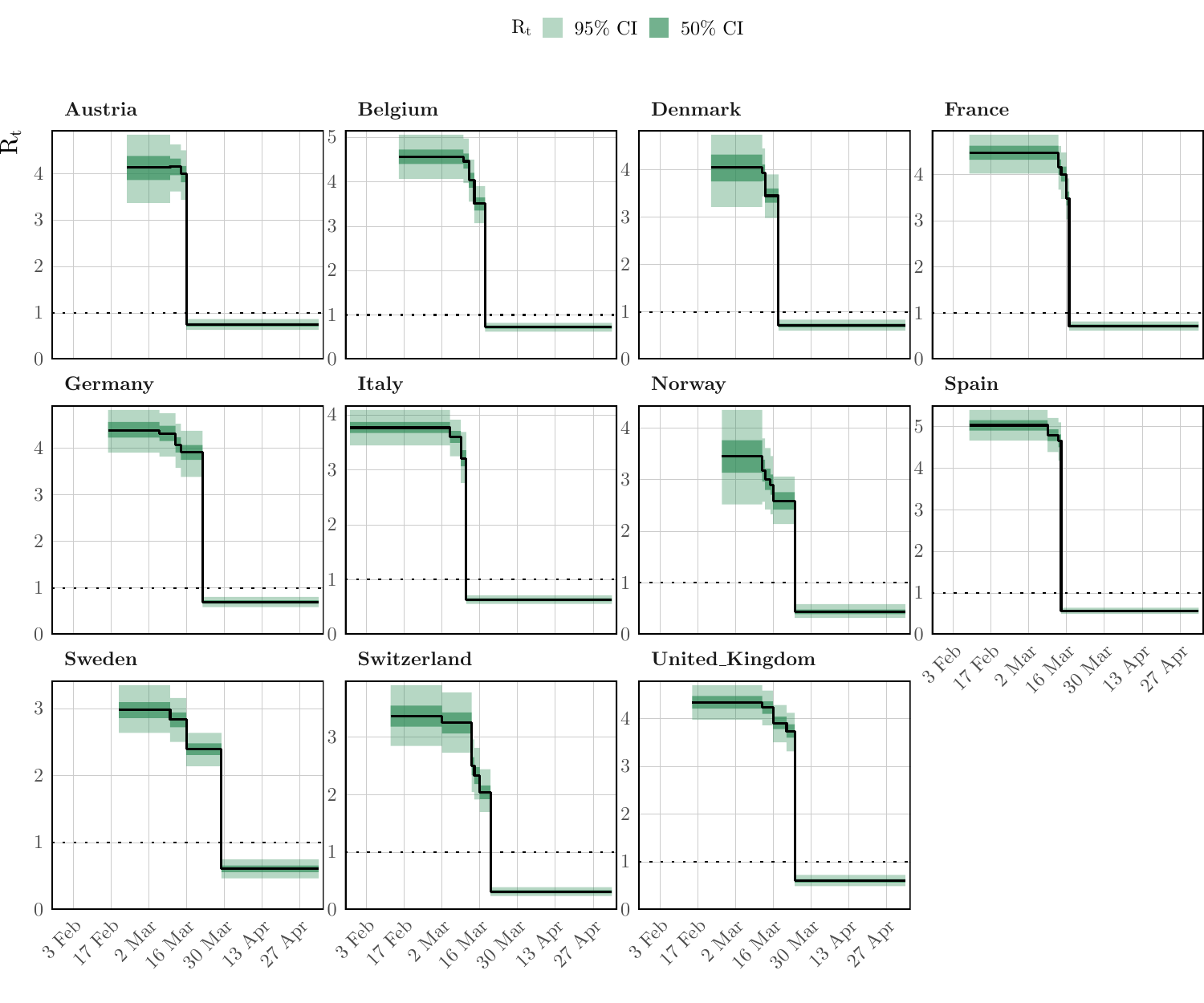}
    \end{adjustbox}
    \caption{\small Inferred reproduction numbers in each country. Credible intervals from the posterior are plotted in shades of green, in addition to the posterior median in black.}
    \label{fig:multilevel-rt-plots}
\end{figure}

Figure \ref{fig:multilevel-obs-plots} suggest that the epidemic was
bought under control in each group considered. Indeed, one would expect
that the posterior distribution for reproduction numbers lies largely
below one in each region. Figure \ref{fig:multilevel-rt-plots} is the
result of \code{plot_rt(fm, step = T, levels = c(50,95))}, and confirms
this.

\hypertarget{effect-sizes}{%
\subsubsection{Effect Sizes}\label{effect-sizes}}

In \pkg{epidemia}, estimated effect sizes can be visualized using the
\code{plot.epimodel} method. This serves a similar purpose to
\code{plot.stanreg} in \pkg{rstanarm}, providing an interface to the
\pkg{bayesplot} package. The models in \pkg{epidemia} often have many
parameters, some of which pertain to a particular part of the model
(i.e.~transmission), and some which pertain to particular groups (i.e.,
country-specific terms). Therefore \code{plot.epimodel} has arguments
\code{par_models}, \code{par_types} and \code{par_groups}, which
restrict the parameters considered to particular parts of the model.

\begin{figure}[ht]
    \centering
    \begin{adjustbox}{width=0.9\textwidth}
    \includegraphics{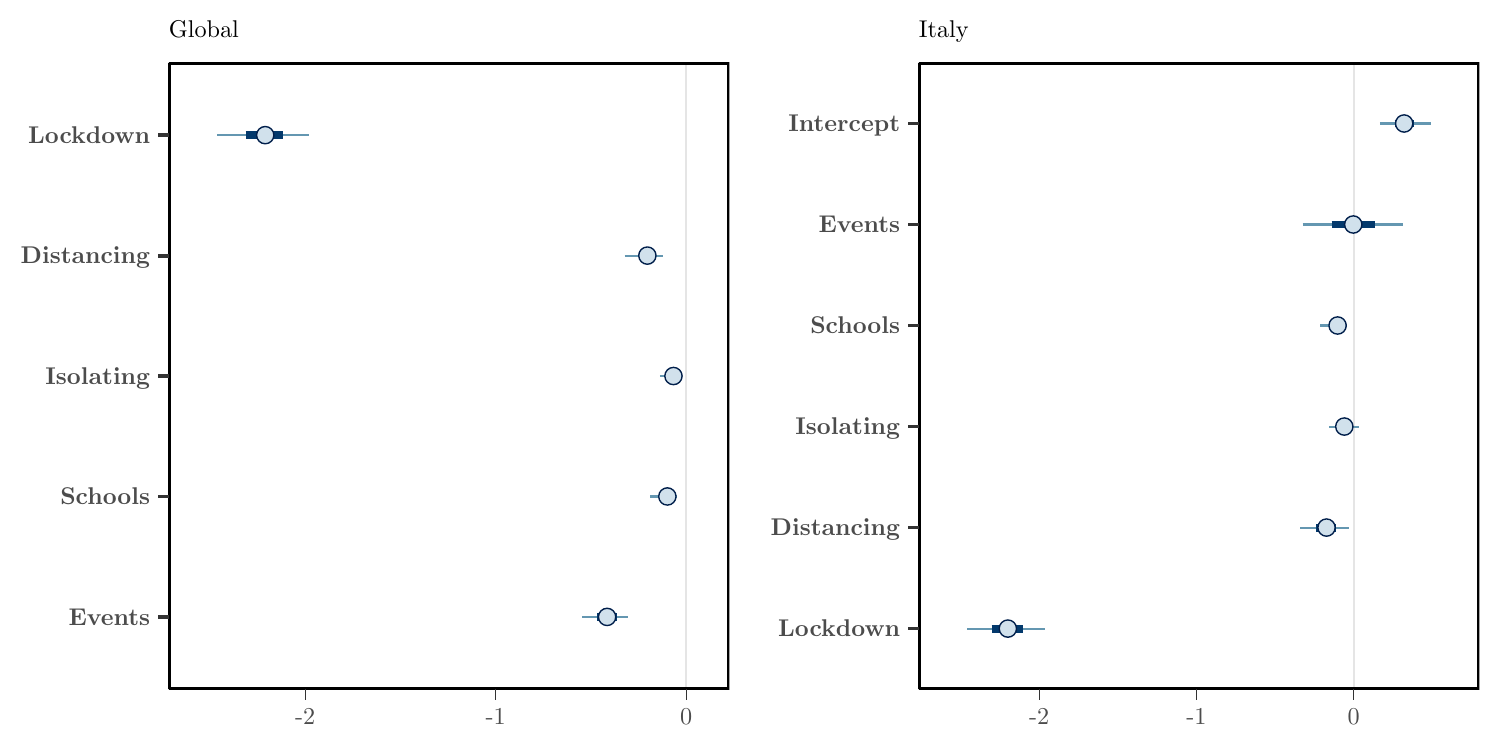}
    \end{adjustbox}
    \caption{\small \textbf{Left}: Global Effect sizes for the five policy measures considered. Right: Effect sizes specific to Italy. The global and country-specific effects may differ because the effects are partially pooled.}
    \label{fig:multilevel-effects-plots}
\end{figure}

As an example, credible intervals for the global coefficients
\(\beta_i\) can be plotted using the command
\code{plot(fm, par_models = "R", par_types = "fixed")}. This leads to
Figure \ref{fig:multilevel-effects-plots}.

Figure \ref{fig:multilevel-effects-plots} shows a large negative
coefficient for lockdown, suggesting that this is on average the most
effective intervention. The effect of banning public events is the next
largest, while the other policy effects appear closer to zero. Note that
the left plot in \ref{fig:multilevel-effects-plots} shows only global
coefficients, and does not show inferred effects in any given country.
To assess the latter, one must instead consider the quantities
\(\beta_i + b^{(m)}_i\). We do this by extracting the underlying draws
using \code{as.matrix.epimodel}, as is done below for Italy.

\begin{CodeChunk}
\begin{CodeInput}
R> beta <- as.matrix(fm, par_models = "R", par_types = "fixed")
R> b <- as.matrix(fm, regex_pars = "^R\\|b", par_groups = "Italy")
R> mat <- cbind(b[,1], beta + b[,2:6])
R> labels <- c("Events", "Schools", "Isolating", "Distancing", "Lockdown")
R> colnames(mat) <- c("Intercept", labels)
\end{CodeInput}
\end{CodeChunk}

Calling \code{bayesplot::mcmc_intervals(mat)} leads to the results shown
in the right panel of Figure \ref{fig:multilevel-effects-plots}.

Figure \ref{fig:multilevel-effects-plots} has relatively narrow
intervals for many of the effect sizes. This appears to be an artifact
of using Variational Bayes. In particular, when repeating this analysis
with full MCMC, we observe that the intervals for all policies other
than lockdown overlap with zero.

\begin{figure}[ht]
    \centering
    \begin{adjustbox}{width=0.9\textwidth}
    \includegraphics{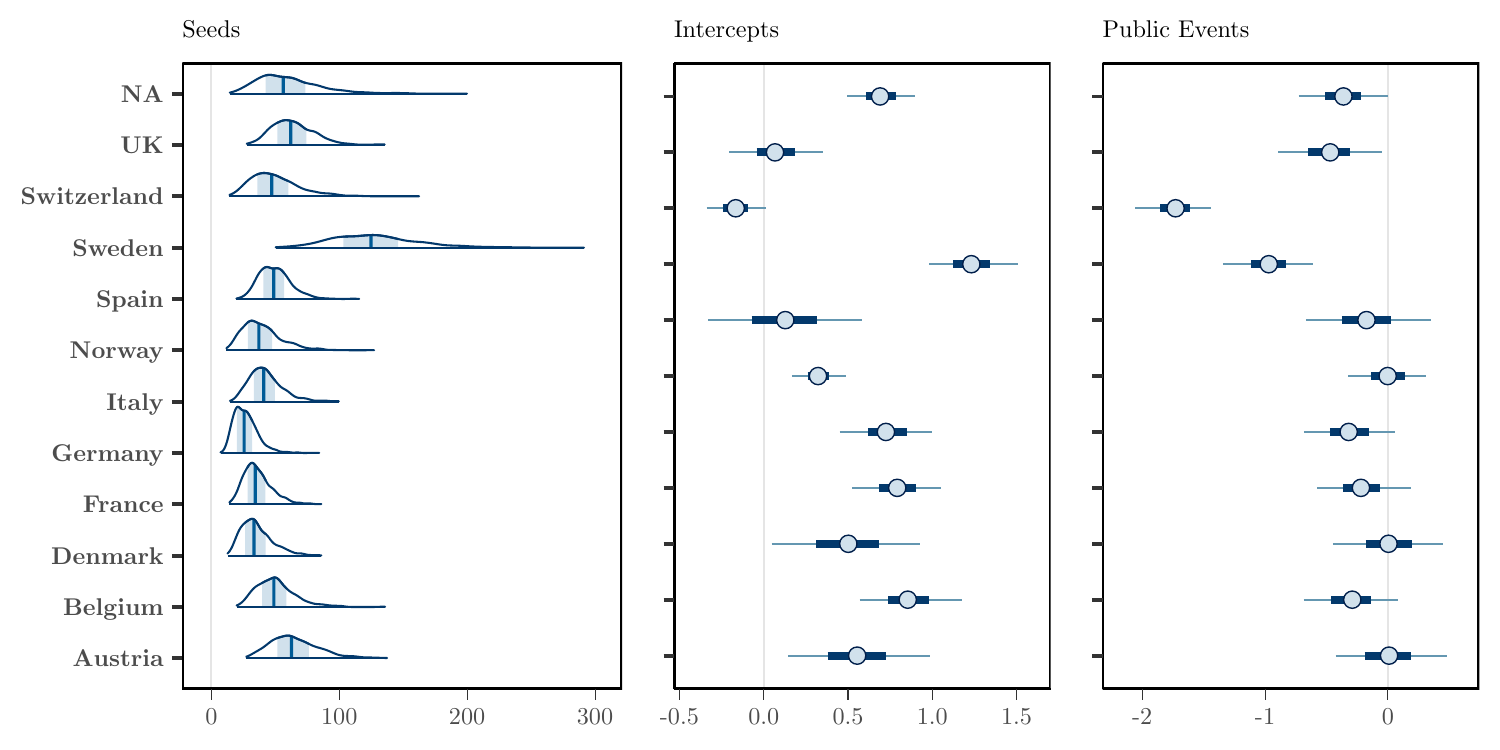}
    \end{adjustbox}
    \caption{\small Left: Inferred daily seeded infections in each country. These have been assumed to occur over a period of 6 days. Middle: Estimated Intercepts in the linear predictor for reproduction numbers. Right: Country-specific effect sizes corresponding to the banning of public events.}
    \label{fig:multilevel-interval-plots}
\end{figure}

Consider now the role of partial pooling in this analysis. Figure
\ref{fig:multilevel-rt-plots} shows that Sweden did enough to reduce
\(R\) below one. However, it did so without a full lockdown. Given the
small effect sizes for other measures, the model must explain Sweden
using the country-specific terms. Figure
\ref{fig:multilevel-interval-plots} shows estimated seeds, intercepts
and the effects of banning public events for each country. Sweden has a
lower intercept than other terms which in turn suggests a lower \(R_0\)
- giving the effects less to do to explain Sweden. There is greater
variability in seeding, because the magnitude of future infections
becomes less sensitive to initial conditions when the rate of growth is
lower. Figure \ref{fig:multilevel-interval-plots} shows that the model
estimates a large negative coefficient for public events in Sweden. This
is significantly larger then the effects for other policies - which are
not reported here. However, the idiosyncrasies relating to Sweden must
be explained in this model by at least one of the covariates, and the
large effect for public policy in Sweden is most probably an artifact of
this. Nonetheless, the use of partial pooling is essential for
explaining difference between countries. If full pooling were used,
effect sizes would be overly influenced by outliers like Sweden. This
argument is made in more detail in \citet{Flaxman2020MA}.

\hypertarget{forecasting}{%
\subsubsection{Forecasting}\label{forecasting}}

Forecasting within \pkg{epidemia} is straightforward, and consists of
constructing a new data frame which is used in place of the original
data frame. This could, for example, change the values of covariates, or
alternatively include new observations in order to check the
out-of-sample performance of the fitted model.

Recall that \code{EuropeCovid2} holds daily death data up until the end
of June 2020, however we only fitted the model up until the
\(5\)\textsuperscript{th} May. The following constructs a data frame
\code{newdata} which contains the additional observations. Note that we
are careful to select the same start dates as in the original data
frame.

\begin{CodeChunk}
\begin{CodeInput}
R> newdata <- EuropeCovid2$data
R> newdata <- filter(newdata, date > date[which(cumsum(deaths) > 10)[1] - 30])
\end{CodeInput}
\end{CodeChunk}

\begin{figure}[ht]
    \centering
    \begin{adjustbox}{width=0.95\textwidth}
    \includegraphics{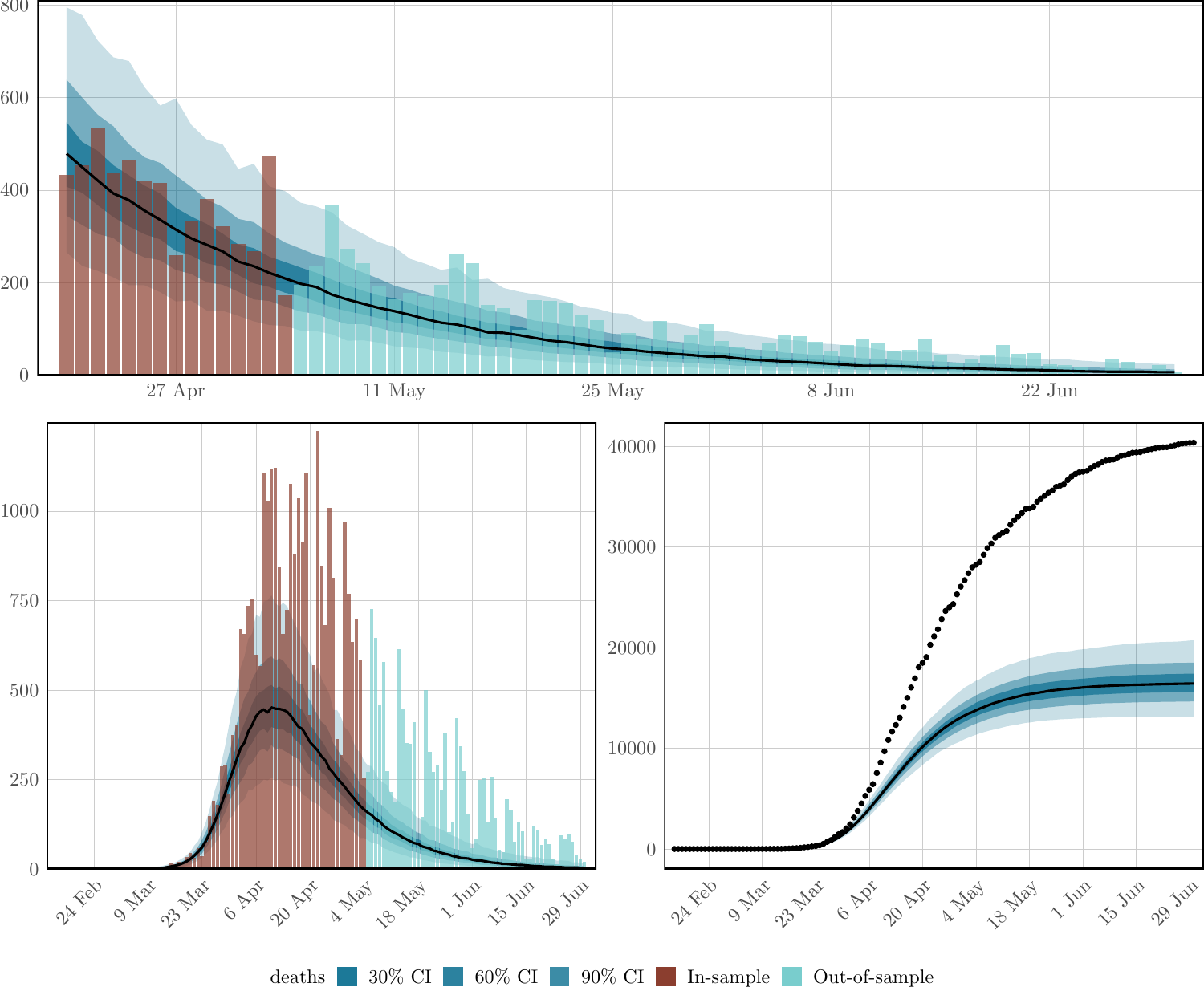}
    \end{adjustbox}
    \caption{\small Forecasts and counterfactual scenarios. All results pertain to the United Kingdom. Top: An out-of-sample forecast for daily deaths. \textbf{Bottom:} Results corresponding to a counterfactual whereby all policies were implemented 3 days earlier. Left: Credible intervals for daily deaths under this scenario. Right: Cumulative deaths. The black dotted line shows observed cumulative deaths.}
    \label{fig:multilevel-forecasting}
\end{figure}

This data frame can be passed to plotting functions \code{plot_rt()},
\code{plot_obs()}, \code{plot_infections()} and
\code{plot_infectious()}. If the raw samples are desired, we can also
pass as an argument to \code{posterior_rt()}, \code{posterior_predict()}
etc. The top panel of Figure \ref{fig:multilevel-forecasting} is the
result of using the command
\code{plot_obs(fm, type = "deaths", newdata = newdata, groups = "Italy")}.
This plots the out of sample observations with credible intervals from
the forecast.

\hypertarget{counterfactuals}{%
\subsubsection{Counterfactuals}\label{counterfactuals}}

Counterfactual scenarios are also easy. Again, one simply has to modify
the data frame used. In this case we shift all policy measures back
three days.

\begin{CodeChunk}
\begin{CodeInput}
R> shift <- function(x, k) c(x[-(1:k)], rep(1,k))
R> days <- 3
R> 
R> newdata <- mutate(newdata,
+   lockdown = shift(lockdown, days),
+   public_events = shift(public_events, days),
+   social_distancing_encouraged = shift(social_distancing_encouraged, days),
+   self_isolating_if_ill = shift(self_isolating_if_ill, days),
+   schools_universities = shift(schools_universities, days)
+ )
\end{CodeInput}
\end{CodeChunk}

The bottom panel of Figure \ref{fig:multilevel-forecasting} visualizes
the counterfactual scenario of all policies being implemented in the UK
three days earlier. Deaths are projected over both the in-sample period,
and the out-of-sample period. The left plot is obtained using
\code{plot_obs(fm, type = "deaths", newdata = newdata, groups = "United_Kingdom")},
while the right plot adds the \texttt{cumulative\ =\ TRUE} argument. We
reiterate that these results are not intended to be fully rigorous: they
are simply there to illustrate usage of \pkg{epidemia}.

\hypertarget{sec:conclusions}{%
\section{Conclusions}\label{sec:conclusions}}

This article has presented \pkg{epidemia}, an \proglang{R} package for
modeling the temporal dynamics of infectious diseases. This is done in a
Bayesian framework, and is regression-oriented, allowing the user
flexibility over model specification. \pkg{epidemia} can be used for a
number of inferential tasks. In particular, the examples of Section
\ref{sec:epi_examples} have demonstrated how to estimate time-varying
reproduction numbers, and to infer the effect of interventions on
disease transmission.

We have not been able to demonstrate all features of \pkg{epidemia}.
Most notably, we have not given examples of applying population
adjustments, using multiple observation vectors, and of starting
modeling at some point after the beginning of an epidemic.

The modeling framework can be extended in a numerous directions.
Currently \(R_t\) can be modeled as a random walk, however additional
autocorrelated processes such as ARMA processes could be considered.
Importations between populations are not currently modeled. This could
be included be adding additional additive terms to the renewal equation
(Equation \ref{eq:renewal2}). More flexible prior distributions for
seeded infections that go beyond the hierarchical model presented here
could be included. Certain epidemiological quantities, such as the
generation distribution are assumed to be known. Uncertainty could be
incorporated by, for example, assigning the generation distribution
could be given a Dirichlet prior. Finally, the question of efficient and
robust fitting of these models is not yet fully resolved. We conjecture
that cleverly selecting starting values for sampling may help prevent
the chains becoming trapped in local modes. This would be an interesting
direction for future research.

\bibliography{covid.bib}

\section*{Appendices}

\hypertarget{sec:priors}{%
\section{Priors on Model Parameters}\label{sec:priors}}

\pkg{epidemia} aims to give the user a high degree of control over
setting prior distributions. It does this by leveraging the
functionality provided by \pkg{rstanarm}, which provides functions
representing a number of different prior families. These include for
example student-t, Laplace, and hierarchical shrinkage families. In this
article, we provide a brief introduction to the available families, and
discuss some important quirks to be aware of when defining priors. We
use the same mathematical notation as in Section
\ref{sec:modeldescription}.

\emph{Please do not rely on the default priors in \pkg{epidemia}.
Although these have been designed to be weakly informative, they are not
guaranteed to be appropriate for your particular model. Please adjust
prior distributions as required.}

Priors must be defined for all parameters in each of the three model
components: transmission, infection, and observations. In the
transmission model, priors must be set for all effects appearing in the
linear predictor \(\eta\). In the infection model, a prior must be set
on \(\tau\), but also on the dispersion parameter \(d\) in the extended
version of the model. In each observational model, priors must be set
for effects defining the multipliers \(\alpha_t\), but also for the
auxiliary parameter for the sampling distribution, \(\phi\).

In general, primitive model parameters can be classified as are either
intercepts, fixed effects, a covariance matrix, an auxiliary parameter,
or the error term in a random walk. We discuss each in turn, in
particular highlighting where they appear in the model, and what
distributions are available for them.

\hypertarget{priors-on-intercepts}{%
\subsection{Priors on Intercepts}\label{priors-on-intercepts}}

Intercepts can appear in the linear predictor \(\eta\) for the
reproduction numbers \(R\) and in the linear predictors for multipliers
\(\alpha\). The prior distribution is specified using an argument
\code{prior_intercept}. This appears in both \code{epirt()} and
\code{epiobs()}. \code{prior_intercept} must be a call to an
\pkg{rstanarm} function that represents a student-t family: i.e.~one of
\code{normal()}, \code{student_t()} or \code{cauchy()} from
\pkg{rstanarm}. \code{prior_intercept} is of course only used if the
formula specifies an intercept. Please note that the interpretation of
\code{prior_intercept} depends on the \code{center} argument to
\code{epirt()} and \code{epiobs()}. Please see Section
\ref{sec:centering} for more details.

\hypertarget{priors-on-regression-coefficients}{%
\subsection{Priors on Regression
Coefficients}\label{priors-on-regression-coefficients}}

In addition to intercepts, the predictors for \(R\) and \(\alpha\) may
also contain fixed effects. In the regression for \(R\) this corresponds
to the parameter vector \(\beta\). The prior distribution is set using
the \code{prior} argument, which, similarly to \code{prior_intercept},
appears in both \code{epirt()} and \code{epiobs()}. Note that this
\emph{does not} set the prior for the group-specific effects \(b\),
which are instead controlled by \code{prior_covariance}.

\code{prior} can be a call to one of \pkg{rstanarm}'s prior functions.
These can be broadly grouped into four families: student-t, hierarchical
shrinkage, Laplace and the product normal family. Note that \emph{all
effects must follow the same family}; for example, it is not possible
for \(\beta_1\) to have a normal prior while \(\beta_2\) has a Cauchy
prior. Nonetheless, different hyperparameters can be set for each
effect.

As an example, suppose the following formula is used to model \(R\),
where \code{cov1} and \code{cov2} are some covariates.

\begin{CodeChunk}
\begin{CodeInput}
R> R(group, date) ~ 1 + cov1 + cov2
\end{CodeInput}
\end{CodeChunk}

Consider the following two prior specifications in the call to
\code{epirt()}.

\begin{itemize}
\item
  \code{prior = rstanarm::normal(location=0,scale=1)} gives a standard
  normal prior to both covariate effects.
\item
  \code{prior = rstanarm::normal(location=c(0,1),scale=c(1,2))} sets
  priors \(\beta_1 \sim N(0,1)\) and \(\beta_2 \sim N(1,2)\), where
  \(\beta_1\) and \(\beta_2\) are the effects for \code{cov1} and
  \code{cov2} respectively. To give different prior locations and or
  scales for each covariate, we simply pass numeric vectors instead of
  scalars.
\end{itemize}

The interpretation of \code{prior} depends on whether covariates are
being centered, and whether automatic scale adjustments are occurring.
Please see Section \ref{sec:caveats} for more details.

\hypertarget{additional-priors}{%
\subsubsection{Additional Priors}\label{additional-priors}}

In addition to \pkg{rstanarm}'s prior functions, \pkg{epidemia} offers
additional prior families for regression coefficients. Currently the
only additional prior available is \code{shifted_gamma}. This represents
a gamma distribution that can be shifted to have support other than on
\([0, \infty)\). Specifically, \begin{equation}
\beta_i \sim \text{Gamma}(\alpha_i, \theta_i) - a_i,
\end{equation} where \(\alpha_i\) and \(\theta_i\) are shape and scale
parameters, and \(a_i\) is a shift. This prior is used in
\citet{Flaxman2020} to model the prior effect of control measures on
Covid-19 transmission. Intuitively, it is unlikely that a measure
designed to reduce transmission rates ends up increasing transmission
significantly. This implies that a symmetric prior may not be
appropriate for these effects: it makes sense to put low mass on large
positive effect sizes. In addition, this prior can help to improve
identifiability when multiple measures occur in quick succession - as is
often the case during the early stages of an epidemic.

\hypertarget{priors-on-auxiliary-parameters}{%
\subsection{Priors on Auxiliary
Parameters}\label{priors-on-auxiliary-parameters}}

Auxiliary parameters can appear in the sampling distributions for
observations. This corresponds to the parameter \(\phi\) introduced in
Section \ref{sec:basic_model}. The interpretation of this parameter
depends on the chosen distribution. The Poisson distribution has no
auxiliary parameter as it is fully defined by its mean. For the negative
binomial distribution (specified by using \code{family = "neg_binom"} in
the call to \code{epiobs()}), \(\phi\) represents the reciprocal
dispersion. An auxiliary parameter \(d\) also exists in the extended
version of the infection model (when using \code{latent = TRUE} in the
call to \code{epiinf()}). See Section \ref{sec:latent} for more
information on this parameter. This represents the \emph{coefficient of
dispersion} of the offspring distribution. Auxiliary parameters are
always non-negative in \pkg{epidemia}.

Priors for auxiliary parameters are set using the \code{prior_aux}
argument in the \code{epiobs()} and \code{epiinf()} modeling functions.
It is not used when \code{family = "poisson"} in the call to
\code{epiobs()} or when \code{latent = FALSE} in the call to
\code{epiinf()}. \code{prior_aux} can be a call to one of
\code{normal()}, \code{student_t()}, \code{cauchy()} or
\code{exponential()} from \pkg{rstanarm}.

\hypertarget{sec:priorscov}{%
\subsection{Priors on Covariance Matrices}\label{sec:priorscov}}

Recall that partial pooling can be used in the regression for \(R_t\).
The partially pooled parameters \(b\) are characterized as zero mean
multivariate normal with an unknown covariance matrix, which must itself
be assigned a prior. The precise model for these parameters is described
in detail in Appendix \ref{sec:partial_pooling}. The prior on the
covariance matrix can be set using the \code{prior_covariance} argument
in \code{epirt()}.

Although the Inverse-Wishart prior is a popular prior for covariance
matrices, it does not cleanly separate shape and scale
\citep{tokuda2011}. A general approach is to decompose the prior on the
covariance matrix into a prior on the correlation matrix and a vector of
variances. This is the approach taken by \pkg{rstanarm}, which has
functions \code{decov()} and \code{lkj()} which represent priors for
covariance matrices. These are also used by \pkg{epidemia} for the same
purpose.

We briefly describe \pkg{rstanarm}'s decov prior, as it applies to
partially pooled parameters in the regression for \(R_t\). Suppose the
formula for \(R_t\) contains a term of the form \code{(expr | factor)},
and that \code{expr} evaluates to a model matrix with \(p\) columns, and
\code{factor} has \(L\) levels. Let \(\theta_l\) denote the \(p\)-vector
of parameters for the \(l\)\textsuperscript{th} group. From Appendix
\ref{sec:partial_pooling} this is modeled as \begin{equation}
  \theta_{l} \sim N(0, \Sigma),
\end{equation} where \(\Sigma\) is a \(p \times p\) covariance matrix.
The decov prior decomposes \(\Sigma\) into a vector of variances
\((\sigma^2_1, \ldots \sigma^2_p)\) and a correlation matrix \(\Omega\),
which is given an LKJ prior. The variance vector is decomposed into the
product of a simplex vector \(s\) and the trace of \(\Omega\), which is
just the sum of the individual variances. Specifically, \begin{equation}
\sigma^2_i = s_i \text{tr}\left(\Sigma\right).
\end{equation} The simplex vector is given a symmetric Dirichlet prior,
while the trace is decomposed into \(tr(\Sigma) = p \kappa^2\), where
\(p\) is the order of the matrix (i.e.~the number of correlated
effects), and \(\kappa\) is a parameter which is assigned a scale
invariant prior; specifically a Gamma with given shape and scale
hyperparameters. When \(p = 1\), for example with \code{(1 | factor)},
the prior simplifies considerably. \(\Sigma\) simply reduces to
\(\kappa^2\), which has a Gamma prior.

\hypertarget{priors-on-random-walks}{%
\subsection{Priors on Random Walks}\label{priors-on-random-walks}}

Section \ref{sec:transmission} described how the linear predictor for
\(R_t\) can include autocorrelation terms. Currently, \pkg{epidemia}
supports random walk terms. The random walk errors are given a zero-mean
normal prior, with an unknown scale. This scale is itself assigned a
half-normal hyperprior with a known scale.

Consider a very simple random walk parameterization of \(R_t\), whereby
\code{formula = R(country, date) ~ rw(prior_scale=0.05)} is used in the
call to \code{epirt()}. Assuming only one population is being
considered, this implies a functional form of \begin{equation*}
R_t = g^{-1}\left(\beta_0 + W_t \right)
\end{equation*} for reproduction numbers. Here \(W_t\) is a random walk
satisfying \(W_t = W_{t-1} + \gamma_t\) for \(t>0\) and with initial
condition \(W_0=0\). Under the prior, the error terms \(\gamma_t\)
follow \(\gamma_t \sim \mathcal{N}(0,\sigma)\) with
\(\sigma \sim \mathcal{N}^{+}(0, 0.05)\).

\hypertarget{sec:caveats}{%
\subsection{Caveats}\label{sec:caveats}}

There are several important caveats to be aware of when using prior
distributions in \pkg{epidemia}.

\hypertarget{sec:centering}{%
\subsubsection{Covariate Centering}\label{sec:centering}}

By default, covariates in the regressions for \(R_t\) and \(\alpha_t\)
are not centered automatically by \pkg{epidemia}. This can, however, be
done by using \code{center = TRUE} in the call to \code{epirt()} and
\code{epiobs()} respectively. It is important to note that if
\code{center = TRUE}, the arguments \code{prior_intercept} and
\code{prior} set the priors on the intercept and coefficients
\textit{after centering the covariates}.

Covariates are not centered automatically because often the intercept
has an intuitive interpretation in the model. For example, if all
covariates are zero at the beginning of the epidemic, then the intercept
can be seen as specifying the initial reproduction number \(R_0\) of the
disease. If \code{center = TRUE}, then the intercept no longer has an
easily intuited interpretation.

\hypertarget{sec:autoscale}{%
\subsubsection{Autoscaling}\label{sec:autoscale}}

\pkg{rstanarm}'s prior functions have an argument called
\code{autoscale}. If \code{autoscale = TRUE}, then \pkg{epidemia}
automatically adjusts the prior scale to account for the scale of the
covariates. This only applies to priors on fixed effects, and not to the
intercepts. \pkg{epidemia} rescales according to the following rules.

\begin{itemize}
\item
  If a predictor has only one unique value, no rescaling occurs.
\item
  If it has two unique values, the original scale is divided by the
  range of the values.
\item
  For more than two unique values, the original scale is divided by the
  standard deviation of the predictor.
\end{itemize}

If you are unsure whether rescaling has occurred, call
\code{prior_summary} on a fitted model object. This gives details on the
original priors specified, and the priors that were actually used after
rescaling.

\hypertarget{sec:partial_pooling}{%
\section{Partial Pooling in epidemia}\label{sec:partial_pooling}}

We describe how to partially pool parameters underlying the reproduction
numbers. This is done using a special operator in the formula passed to
\code{epirt()}. If you have previously used any of the \pkg{lme4},
\pkg{nlmer}, \pkg{gamm4}, \pkg{glmer} or \pkg{rstanarm} packages then
this syntax will be familiar.

A general \proglang{R} formula is written as \code{y ~ model}, where
\code{y} is the response that is modeled as some function of the linear
predictor which is symbolically represented by \code{model}.
\code{model} is made up of a series of terms separated by \code{+}. In
\pkg{epidemia}, as in many other packages, parameters can be partially
pooled by using terms of the form \code{(expr | factor)}, where both
\code{expr} and \code{factor} are \proglang{R} expressions. \code{expr}
is a standard linear model (i.e.~treated the same as \code{model}), and
is parsed to produce a model matrix. The syntax \code{(expr | factor)}
makes explicit that columns in this model matrix have separate effects
for different levels of the factor variable.

Of course, separate effects can also be specified using the standard
interaction operator \code{:}. This however corresponds to \emph{no
pooling}, in that parameters at different levels are given separate
priors. The \code{|} operator, on the other hand, ensures that effects
for different levels are given a common prior. This common prior itself
has parameters which are given hyperpriors. This allows information to
be shared between different levels of the factor. To be concrete,
suppose that the model matrix parsed from \code{expr} has \(p\) columns,
and that \code{factor} has \(L\) levels. The \(p\)-dimensional parameter
vector for the \(l\)\textsuperscript{th} group can be denoted by
\(\theta_l\). In \pkg{epidemia}, this vector is modeled as multivariate
normal with an unknown covariance matrix. Specifically, \begin{equation}
  \theta_{l} \sim N(0, \Sigma),
\end{equation} where the covariance \(\Sigma\) is given a prior.
\pkg{epidemia} offers the same priors for covariance matrices as
\pkg{rstanarm}; in particular the \code{decov()} and \code{lkj()} priors
from \pkg{rstanarm} can be used. Note that \(\Sigma\) is not assumed
diagonal, i.e.~ the effects within each level may be correlated.

If independence is desired for parameters in \(\theta_l\), we can simply
replace \code{(expr | factor)} with \code{(expr || factor)}. This latter
term effectively expands into \(p\) terms of the form
\code{(expr_1 | factor)}, \(\ldots\), \code{(expr_p | factor)}, where
\code{expr_1} produces the first column of the model matrix given by
\code{expr}, and so on. From the above discussion, the effects are
independent across terms, and essentially \(\Sigma\) is replaced by
\(p\) one-dimensional covariance matrices (i.e.~variances).

\hypertarget{example-formulas.}{%
\subsection{Example Formulas.}\label{example-formulas.}}

The easiest way to become familiar with how the \code{|} operator works
is to see a multitude of examples. Here, we give many examples, their
interpretations, and where possible we compare the models to the no
pooling and full pooling equivalents. For a comprehensive reference on
mixed model formulas, please see \citet{bates_2015}.

There are many possible ways to specify intercepts. Table
\ref{tab:intercept-specs} demonstrates some of these, including fully
pooled, partially pooled and unpooled. Effects may also be partially
pooled. This is shown in Table \ref{tab:cov-specs}.

\begin{CodeChunk}
\begin{table}[!h]

\caption{\label{tab:intercept-specs}\small Different intercept specifications. The intercept often has an interpretation as setting $R_0$ in each region. The left hand side of each formula is assumed to take the form \code{R(region, date)}.}
\centering
\begin{tabular}[t]{>{\raggedright\arraybackslash}p{15em}>{\raggedright\arraybackslash}p{20em}}
\toprule
Formula R.H.S. & Interpretation\\
\midrule
\textbf{\code{1 + ...}} & Full pooling, common intercept for all regions.\\
\textbf{\code{region + ...}} & Separate intercepts for each region, not pooled.\\
\textbf{\code{(1 | region) + ...}} & Separate intercepts for each region which are partially pooled.\\
\textbf{\code{(1 | continent) + ...}} & Separate intercepts based on a factor other than \code{region}, partially pooled.\\
\bottomrule
\end{tabular}
\end{table}

\end{CodeChunk}

\begin{CodeChunk}
\begin{table}[!h]

\caption{\label{tab:cov-specs} \small Different covariate specifications. Here NPI refers to some non-pharamceutical intervention. The left hand side of each formula is assumed to take the form \code{R(region, date)}.}
\centering
\begin{tabular}[t]{>{\raggedright\arraybackslash}p{15em}>{\raggedright\arraybackslash}p{20em}}
\toprule
Formula R.H.S. & Interpretation\\
\midrule
\textbf{\code{1 + npi + ...}} & Full pooling. Effect of NPI the same across all regions.\\
\textbf{\code{1 + npi:region + ...}} & No pooling. Separate effect in each region.\\
\textbf{\code{1 + (0 + npi|region) + ...}} & Partial pooling. Separate effects in each region.\\
\textbf{\code{1 + (npi|region) + ...}} & Right hand side expands to \code{1 + (1 + npi|region)}, and so both the intercept and effect are partially pooled.\\
\bottomrule
\end{tabular}
\end{table}

\end{CodeChunk}

The final example in Table \ref{tab:cov-specs} shows that it is
important to remember that to parse the term \code{(expr | factor)},
\code{epim()} first parses \code{expr} into a model matrix in the same
way as functions like \code{lm()} and \code{glm()} parse models. In this
case, the intercept term is implicit. Therefore, if this is to be
avoided, we must explicitly use either \code{(0 + npi | region)} or
\code{(-1 + npi | region)}.

\hypertarget{independent-effects}{%
\subsubsection{Independent Effects}\label{independent-effects}}

By default, the vector of partially pooled intercepts and slopes for
each region are correlated. The \code{||} operator can be used to
specify independence. For example, consider a formula of the form

\begin{CodeChunk}
\begin{CodeInput}
R> R(region, date) ~ npi + (npi || region) + ...
\end{CodeInput}
\end{CodeChunk}

The right hand side expands to
\code{1 + npi + (1 | region) + (npi | region) + ...}. Separate
intercepts and effects for each region which are partially pooled. The
intercept and NPI effect are assumed independent within regions.

\hypertarget{nested-groupings}{%
\subsubsection{Nested Groupings}\label{nested-groupings}}

Often groupings that are nested. For example, suppose we wish to model
an epidemic at quite a fine scale, say at the level of local districts.
Often there will be little data for any given district, and so no
pooling will give highly variable estimates of reproduction numbers.
Nonetheless, pooling at a broad scale, say at the country level may hide
region specific variations.

If we have another variable, say \code{county}, which denotes the county
to which each district belongs, we can in theory use a formula of the
form

\begin{CodeChunk}
\begin{CodeInput}
R> R(district, date) ~ (1 | county / district) + ...
\end{CodeInput}
\end{CodeChunk}

The right hand side expands to
\code{(1 | county) + (1 | county:district)}. There is a county level
intercept, which is partially pooled across different counties. There
are also district intercepts which are partially pooled \emph{within}
each county.

\hypertarget{sec:modelschematic}{%
\section{Model Schematic}\label{sec:modelschematic}}

We provide schematics for different parts of the model introduced in
Section \ref{sec:modeldescription}. These are useful because they
clarify how different model objects, including data and parameters, are
related to one another.

\begin{figure}[ht]
    \centering
    \begin{tikzpicture}
    \foreach \x in {0,...,5}
        \foreach \y in {0,...,9}
        {
            \node[dummy] (\x\y) at (1.8*\x,1.8*\y) {};
        }
        
    \node[ minimum height = 16.4cm, minimum width=6.8cm, fill = anti-flashwhite!50] at (1.8, 8.8) (obsmodel1) {};
    \node[above of = obsmodel1, node distance=8.4cm, minimum width = 6.8cm, minimum height = 0.6cm, fill=arsenic, text=white] (obs1lab) {First Observation Model};
    \node[rectangle, rounded corners=0.2cm, minimum height = 9cm, minimum width = 6.4cm, fill=aquamarine!50] (multmodel) at (1.8,5.25) {};
    \node[] (multmodel) at (3.4,10) {\textbf{Multiplier Model}};

    \node[tparam, draw=amber] (infections) at (30) {Infections \\ $\{i_s, s <t\}$};
    \node[tparam] (ascertainment1) at (15) {Multiplier \\ $\alpha^{(1)}_t$};
    \node[quant] (link1) at (04) {Link \\ $g_1$};
    \node[tparam] (predictor1) at (13) {Predictor \\ $\eta^{(1)}_t$};
    \node[minimum height = 1.7cm, minimum width = 2.6cm] (data1) at (01) {};
    \node[quant, minimum height = 0.6cm, minimum width = 2.4cm] (fixed_data1) at (0,2.2) {Fixed: \small $x^{(1)}_t$};
    \node[quant, minimum height = 0.6cm, minimum width = 2.4cm]
    (autocor_data1) at (0,1.4) {Autocor: \small $q^{(1)}_t$};
    \node[minimum height = 1.7cm, minimum width = 2.6cm] (param1) at (21) {};
    \node[param, minimum height = 0.6cm, minimum width = 2.4cm] (fixed_param1) at (3.6,2.2) {Fixed: $\beta_1$};
    \node[param, minimum height = 0.6cm, minimum width = 2.4cm](autocor_param1) at (3.6,1.4) {Autocor: $\gamma_1$};
    \node[circle, inner sep=2pt, align=center] (addmodels) at (56) {More Observation \\ Models};
    \node[quant] (offset1) at (02) {Offset: $o_t$};
    \node (label) at (3.6, 2.9) {\textbf{Effects}};
    \node (label) at (0, 2.9) {\textbf{Data}};

\node[quant] (i2o1) at (06) {Inf $\Rightarrow$ Obs \\ $\{\pi^{(1)}_k\}$};
\node[tparam] (eobs1) at (17) {Expected Obs  \\ $y^{(1)}_t$};
\node[param] (aux1) at (28) {Auxiliary \\ $\phi_1$};
\node[quant] (dist1) at (08) {Family \\ $p_1(\cdot, y^{(1)}_t, \phi_1)$};
\node[obs] (obs1) at (19) {Observations \\ $Y^{(1)}_t$};
\path[line,-] (infections.north) to (36.south);
\path[line,-] (36.south) to[bend right=45] (26.east);
\path[line,-] (36.south) to[bend left=45] (46.west);
\path[line,-] (26.east) to (26.west);
\path[line,-] (26.west) to[bend left=45] (16.north);
\path[line,-] (i2o1.east) to[bend right=45] (16.north);
\path[line,-] (aux1.west) to[bend left=45] (18.north);
\path[line,-] (dist1.east) to[bend right=45] (18.north);
\path[line] (ascertainment1.north) to (eobs1.south);
\path[line] (eobs1.north) to (obs1.south);
\path[line,-] (data1.east) to[bend right=45] (11.north);
\path[line,-] (param1.west) to[bend left=45] (11.north);
\path[line] (11.north) to (predictor1.south);
\path[line,-] (link1.east) to[bend right=45] (14.north);
\path[line] (predictor1.north) to (ascertainment1.south);
\path[line, dotted, -] (46.west) to (addmodels.west);
\path[line,-] (offset1.east) to[bend right=45] (12.north);
\end{tikzpicture}
    \caption {\small A schematic for observational models. Only one observational model is shown here, however the figure makes clear that additional models may be included. The model for the multiplier $\alpha_t$ is shown in the shaded green region. This is very similar in form to the transmission model shown in Figure \ref{fig:inf_basic}. Infections shown at the bottom may be directly from either the basic infection model, or from an extended model (as described in Section \ref{sec:modeldescription}.}
    \label{fig:obs_model}
\end{figure}
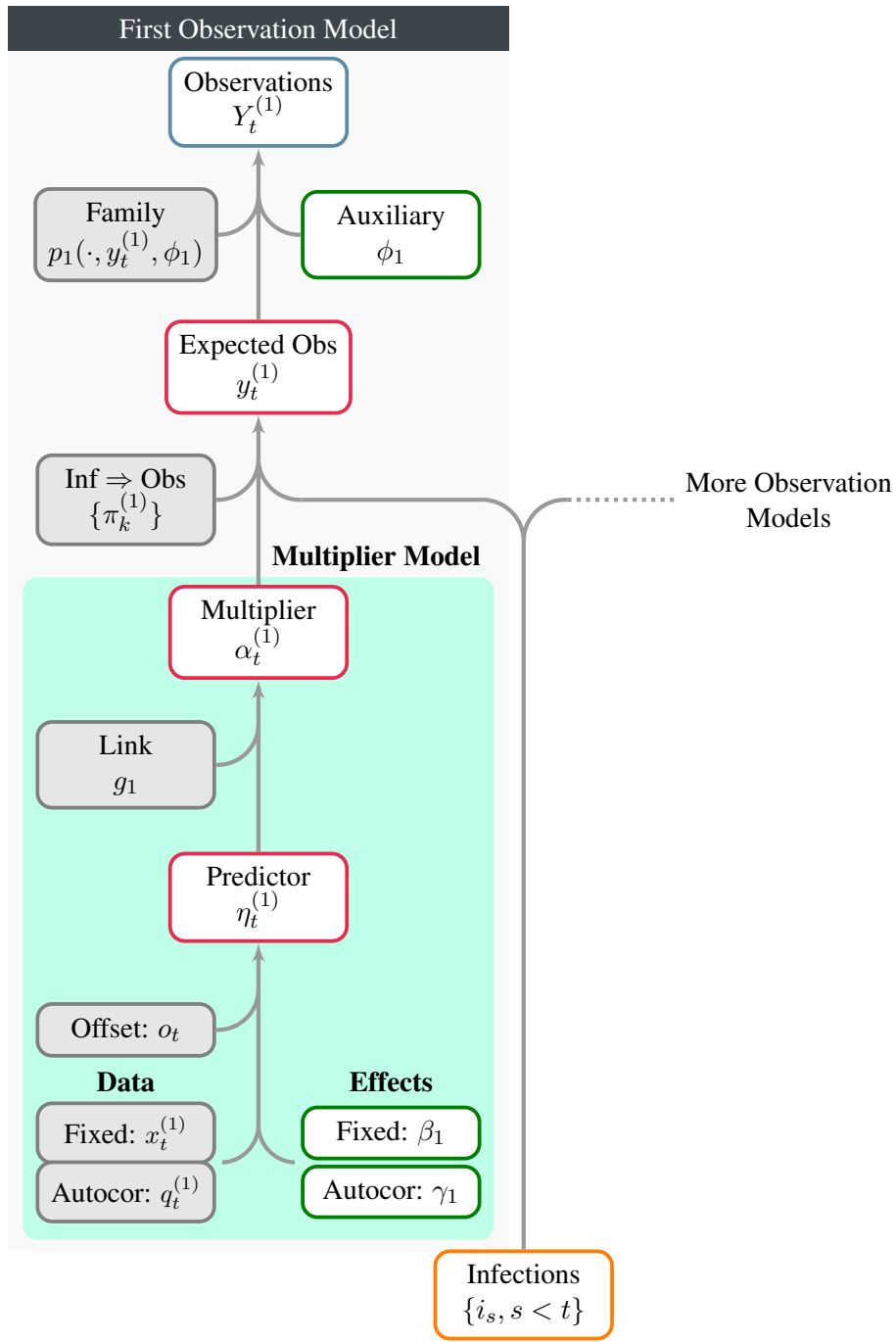

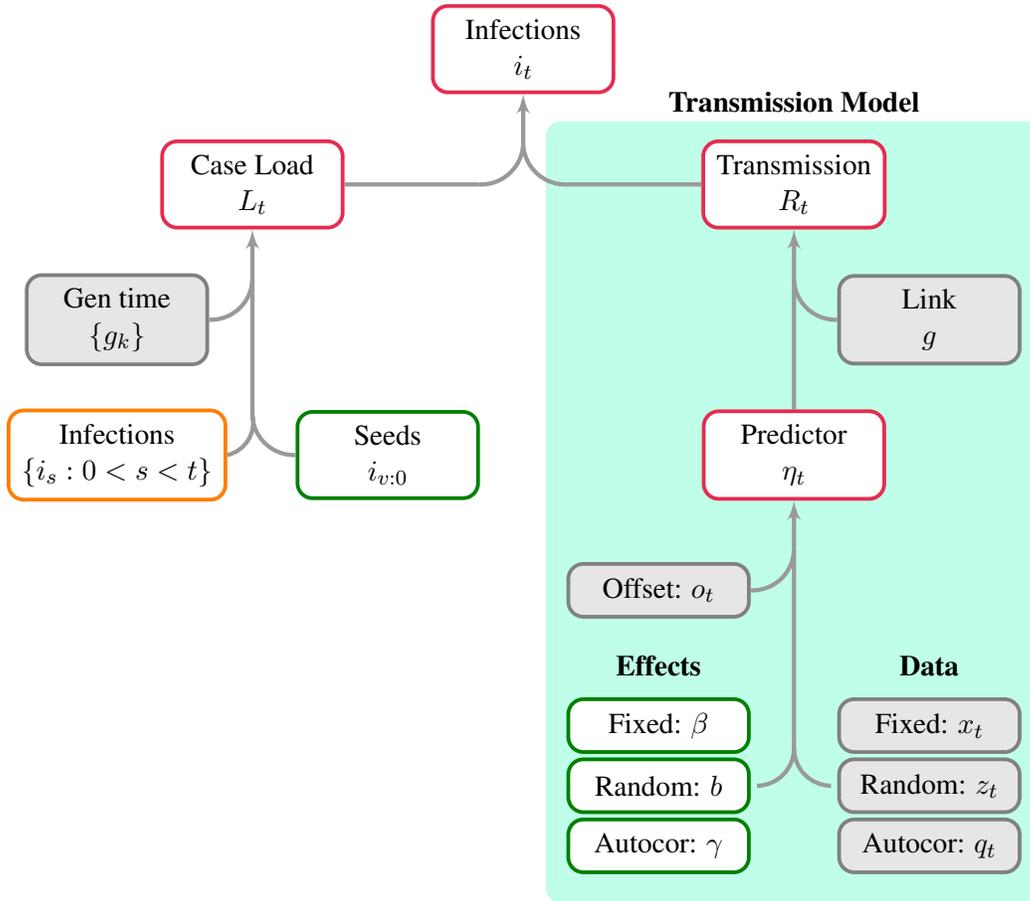
\begin{figure}[ht]
    \centering
    \begin{tikzpicture}
        \foreach \x in {0,...,6}
            \foreach \y in {0,...,5}
            {
                \node[dummy] (\x\y) at (1.8*\x,1.8*\y) {};
            }
            \node[rectangle, rounded corners=.2cm,fill=aquamarine!50, minimum height = 10.4cm, minimum width = 6.6cm] (transmodel) at (9,2.84) {};
            \node (transmodellabel) at (9, 8.3) {\textbf{Transmission Model}};
            \node[tparam, draw=amber] (pastinf) at (02) {Infections \\ $\{i_s: 0 < s <t\}$};
            \node[quant] (gen) at (03) {Gen time \\ $\{g_k\}$};
            \node[param] (seeds) at (22) {Seeds \\ $i_{v:0}$};
            \node[tparam] (load) at (14) {Case Load \\ $L_t$};
            \node[tparam] (inf) at (35) {Infections \\ $i_t$};
            \node[tparam] (transmission) at (54) {Transmission \\ $R_t$};
            \node[tparam] (predictor) at (52) {Predictor \\ $\eta_t$};
            \node[quant] (link) at (63) {Link \\ $g$};
            \node[minimum height = 2.6cm, minimum width = 2.6cm] (effects) at (7.2, -0.8) {};
            \node (label) at (7.2, 0.8) {\textbf{Effects}};
             \node (label) at (10.8, 0.8) {\textbf{Data}};
            \node[param] (fixed) at (7.2,0) {Fixed: $\beta$};
            \node[param] (random) at (7.2,-0.8) {Random: $b$};
            \node[param] (autocor) at (7.2,-1.6) {Autocor: $\gamma$};
            \node[minimum height = 2.6cm, minimum width = 2.6cm] (data) at (10.8, -0.8) {};
            \node[quant] (fixed) at (10.8,0) {Fixed: $x_t$};
            \node[quant] (random) at (10.8,-0.8) {Random: $z_t$};
            \node[quant] (autocor) at (10.8,-1.6) {Autocor: $q_t$};
            \node[quant] (offset) at (41) {Offset: $o_t$};
            \path[line,-] (pastinf.east) to[bend right=45] (12.north);
            \path[line] (12.north) to (load.south);
            \path[line,-] (seeds.west) to[bend left=45] (12.north);
            \path[line,-] (gen.east) to[bend right=45] (13.north);
            \path[line,-] (effects.east) to[bend right=45] (9, -0.3);
            \path[line,-] (data.west) to[bend left=45] (9, -0.3);
            \path[line] (9,-0.3) to (predictor.south);
            \path[line,-] (offset.east) to[bend right=45] (51.north);
            \path[line,-] (link.west) to[bend left=45] (53.north);
            \path[line] (predictor.north) to (transmission.south);
            \path[line,-] (transmission.west) to (44.west);
            \path[line,-] (load.east) to (24.east);
            \path[line,-] (44.west) to[bend left=45] (34.north);
            \path[line,-] (24.east) to[bend right=45] (34.north);
            \path[line] (34.north) to (inf.south);
    \end{tikzpicture}
    \caption {\small A schematic showing both the basic infection model and the transmission model (the green region). Here infections are a transformed parameter, and are recursively linked to previous infections. The model for $R_t$ is similar to a GLM, however autocorrelation terms can be included. $\eta_t$ is the predictor for the reproduction number at time $t$, and is one element of the predictor $\eta$ introduced in Section \ref{sec:modeldescription}}
    \label{fig:inf_basic}
\end{figure}

\begin{figure}
    \centering
    \begin{subfigure}{0.4\textwidth}
    \resizebox{\textwidth}{!}{
    \begin{tikzpicture}
        \foreach \x in {0,...,2}
            \foreach \y in {0,...,2}
            {
                \node[dummy] (\x\y) at (1.8*\x,1.8*\y) {};
            }
            \node[tparam] (inf) at (10) {Infections \\ $i_t$};
            \node[param] (cov) at (01) {CoV \\ $d$};
            \node[quant] (dist) at (21) {Distribution \\ $p(\cdot; d)$};
            \node[param] (infparam) at (12) {Infections \\ $i_t$};
            \path[line,-] (cov.east) to[bend right=45] (11.north);
            \path[line,-] (dist.west) to[bend left=45] (11.north);
            \path[line] (inf.north) to (infparam.south);
\end{tikzpicture}
    }
    \end{subfigure}
    \begin{subfigure}{0.5\textwidth}
    \resizebox{\textwidth}{!}{
    \begin{tikzpicture}
\foreach \x in {0,...,4}
    \foreach \y in {0,...,3}
    {
        \node[dummy] (\x\y) at (1.8*\x, 1.8*\y) {};
    }
    \node[quant] (vacc) at (00) {Vaccinations \\ $V_t-V_{t-1}$};
    \node[tparam] (immunepast) at (20) {Immune \\ $Q_{t-1}$};
    \node[tparam, draw=amber] (infunadj) at (40) {Infections \\ $i_{t}$};
    \node[tparam] (effvacc) at (02) {Effective \\ $v_t$};
    \node[tparam] (immune) at (22) {Immune \\ $Q_{t}$};
    \node[tparam] (inf) at (42) {Infections \\ $i_{t}$};
    \node[quant] (pop) at (23) {Population \\ $P$};
    \path[line,-] (41.south) to[bend right=45] (31.east);
    \path[line,-] (21.south) to[bend right=45] (11.east);
    \path[line,-] (21.north) to[bend right=45] (31.west);
    \path[line,-] (01.north) to[bend right=45] (11.west);
    \path[line,-] (11.west) to (31.east);
    \path[line] (vacc.north) to (effvacc.south);
    \path[line] (immunepast.north) to (immune.south);
    \path[line] (infunadj.north) to (inf.south);
    \path[line] (effvacc.east) to (immune.west);
    \path[line] (immune.east) to (inf.west);
    \path[line,-] (pop.west) to (13.west);
    \path[line,-] (13.west) to[bend right=45] (03.south);
    \path[line,-] (pop.east) to (33.east);
    \path[line,-] (33.east) to[bend left=45] (43.south);
    \path[line] (03.south) to (effvacc.north);
    \path[line] (43.south) to (inf.north);
\end{tikzpicture}
    }
    \end{subfigure}
    \caption{\small Possible extensions to the infection process. \textbf{Left} corresponds to the extension of Section \ref{sec:latent}, while \textbf{right} shows the extension of Section \ref{sec:epi_popadjust}. The population adjustment, shown in the right figure, may be applied to either the infections shown at the bottom of the left figure (basic model), or those at the top of the left figure.}
    \label{fig:inf_extensions}
\end{figure}
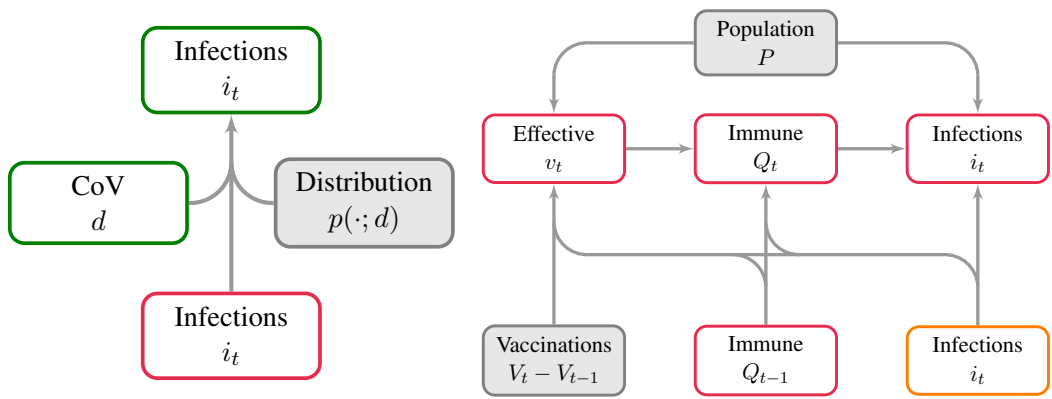

Figures \ref{fig:obs_model} illustrates a complete observational model,
and in particular details the model for multipliers \(\alpha_t\). Figure
\ref{fig:inf_basic} presents the basic infection model, and also shows
the GLM-style model for reproduction numbers \(R_t\). Finally Figure
\ref{fig:inf_extensions} shows extensions of the basic infection model,
including treating latent infections as parameters and including
population adjustments.

All mathematical notation shown in the figures corresponds to that used
in Section \ref{sec:modeldescription}. Each node is outlined in a color
corresponding to the type of object considered. These are interpreted as
follows.

\begin{itemize}
\item
  \textbf{Grey}: A user provided object or quantity that is assumed to
  be known.
\item
  \textbf{Green}: A model parameter that is, generally speaking,
  directly sampled. Occasionally \pkg{epidemia} will sample a
  transformation of this parameter for efficiency purposes.
\item
  \textbf{Red}: A transformed parameter. This is a quantity that is a
  deterministic function of other model parameters.
\item
  \textbf{Orange}: A quantity that is either a parameter or transformed
  parameter, depending on the context.
\item
  \textbf{Blue}: An observation.
\end{itemize}

\end{document}